\documentclass[preprint,12pt]{elsarticle}

\usepackage{graphicx}

\usepackage{amssymb}
\usepackage{amsthm}

\usepackage{color}

\usepackage{lineno}

\usepackage{diagbox} 
\usepackage{multicol} 
\usepackage{amsmath}

\newcounter{bla}

\journal{Computer Physics Communications}

\begin{document}

\begin{frontmatter}



\title{PANNA: Properties from Artificial Neural Network Architectures}


\author[a]{Ruggero Lot}
\author[a]{Franco Pellegrini}
\author[a]{Yusuf Shaidu}
\author[a]{Emine K\"u\c{c}\"ukbenli\corref{author}}

\cortext[author] {Corresponding author.\\\textit{E-mail address:} kucukben@g.harvard.edu \\
Current affiliation: John A. Paulson School of Engineering and Applied Sciences, Harvard University, Cambridge Massachusetts 02138, USA}
\address[a]{SISSA, Via Bonomea 265, I-34136 Trieste, Italy}

\begin{abstract}
Prediction of material properties from first principles is often a computationally expensive task. Recently, artificial neural networks and other machine learning approaches have been successfully employed to obtain accurate models at a low computational cost by leveraging existing example data.
Here, we present a software package ``Properties from Artificial Neural Network Architectures'' (PANNA) that provides a comprehensive toolkit for creating neural network models for atomistic systems. Besides the core routines for neural network training, it includes data parser, descriptor builder and force-field generator suitable for integration within molecular dynamics packages. PANNA offers a variety of activation and cost functions, regularization methods, as well as the possibility of using fully-connected networks with custom size for each atomic species. PANNA benefits from the optimization and hardware-flexibility of the underlying TensorFlow engine which allows it to be used on multiple CPU/GPU/TPU systems, making it possible to develop and optimize neural network models based on large datasets.


\end{abstract}


\end{frontmatter}

{\bf PROGRAM SUMMARY}

\begin{small}
\noindent
{\em Program Title:} PANNA -- Properties from Artificial Neural Network Architectures\\
{\em Licensing provisions:}  MIT\\
{\em Programming language:} Python, C++\\
{\em Nature of problem:}\\
A workflow for machine learning atomistic properties and interatomic potentials using neural networks. 
\\
{\em Solution method:}\\
This package first transforms the user supplied data into pairs of precomputed input and target output for the neural network model. The data are then packed to enable efficient reading. A user-friendly interface to TensorFlow~[1] is provided to instantiate and train neural network models with varying architectures and training schedules. The training can be monitored and validated with the provided tools. The derivative of the target output with respect to the input can also be used jointly in training, e.g. in the case of energy and force training. The interface with molecular dynamics codes such as LAMMPS~[2] allows the neural network model to be used as an interatomic potential.
\\
{\em Unusual features:}\\
The package allows different network architectures to be used for each atomic species, with different trainability setting for each network layer. It provides tools of exchanging weights between atomic species, and provides the option of building a Radial Basis Function network. The software is parallelized to take advantage of hardware architectures with multiple CPU/GPU/TPUs.\\
{\em Additional comments:}\\
The underlying neural network training engine, TensorFlow, is a prerequisite of PANNA. While there is a special LAMMPS integration performed via a patch distributed within PANNA, the network potentials can be deposited into OpenKIM~[3] database and can be used with a wide range of molecular dynamics codes.\\

\end{small}

\clearpage


\section{Introduction}

In recent years machine learning has gained increasing popularity in material science and chemical physics due to several potentially high-impact applications: 
Electronic properties such as atomization energy, polarizability, infrared spectra and excitation energies have been calculated via machine learning with satisfactory accuracy for small organic molecules~\cite{Montavon2013, Hirn2017, Yao2018TensorMol} and molecular crystals~\cite{Musil2018}. New methodologies have been developed for inorganic crystals in order to successfully predict electronic structure properties such as the density of states~\cite{Chandrasekaran2019, Schutt2014}, Debye temperature~\cite{Isayev2017}, band gap of inorganic crystals~\cite{Rajan2018}.
In these studies, various approaches have been employed to represent atomic systems to be used within machine learning algorithms. Some examples from literature are vectors with respect to local frames \cite{Wang2018178}, symmetry-based descriptors~\cite{BP07,Bartok2013}, graphs~\cite{Xie2018}, matrices~\cite{Rupp2012}, list of bonds~\cite{Hansen2015}, chemical formulas~\cite{Legrain2017} or molecular structures~\cite{Tsubaki2018}.
These respresentations are often coupled with an appropriate machine learning model such as feedforward neural networks~\cite{Blank1995}, convolutional networks~\cite{Schutt2018}, Gaussian processes~\cite{Bartok2010}.
It has been demonstrated that a successful pairing of representation and machine learning model can be found to predict local properties such as atomic charges~\cite{Nebgen2018} or electronic density~\cite{Brockherde2017,Grisafi2018,Sinitskiy2018}. 
Forces on atoms, a local property that is trivially linked to the global total energy via gradients, have been calculated both analytically \cite{Unke2019} and numerically as a target of the machine learnt model \cite{CHMIELA201938}.
Fast and accurate interatomic force fields have been developed based on neural networks~\cite{Behler2011,Zhang2018,Huang2019} and Gaussian processes~\cite{Li2015,Glielmo2017, Rowe2018}.

Despite the large amount of publications that demonstrate the success of machine-learnt force fields in proof-of-concept scale, there have been fewer investigations where they have been used in production scale and found to successfully complement the existing force field database, e.g. in the case of disordered materials~\cite{Artrith2014}, catalytic surfaces~\cite{Kolsbjerg2018}, nanoclusters~\cite{Zeni2018}.
Although available datasets~\cite{Mounet2018,Jain2016,Gossett2018} and innovative machine learning methods~\cite{Gilmer2017} increase, considerations remain in adopting these methods for applications beyond the proof-of-concept scale. In particular, a high performance neural network training engine requires substantial effort to implement. Furthermore, the accuracy of a network model depends on the successful optimization of several parameters of the model, and introducing such hyperparameter changes in a neural network program may require extensive coding overhead for non-experts. For the resulting models to be used in realistic research questions, integration of these models into popular massively parallel molecular dynamics (MD) packages is a must. Hence for neural network potentials to find their place in the state-of-the art production arsenal of application scientists, tools that offer accuracy and performance for a wide range of applications are needed. Likewise, challenging real world applications can reveal the areas of improvement for obtaining better neural network models. One way forward is an open source package that gives a simple way to develop, test and use different neural network models efficiently for large amounts of datasets, integrated with MD and other material science simulation packages, without re-inventing the infrastructure necessary for atomistic machine learning training each time. To match this need we have built PANNA (Properties from Artificial Neural Network Architectures), a Python package based on TensorFlow that simplifies the process of training, testing and using neural network models in atomistic calculations.

PANNA includes tools to parse and convert ab initio simulation files into neural network (NN) inputs, the training and monitoring of NN models, conversion of the network into interatomic potential format that can be used in MD simulations. The implementation supports fully-connected network architecture, a variety of input and output formats, and it can be run on large parallel CPU/GPU/TPU systems.

In the following sections we describe the details of the implementation, in particular: in Section~\ref{s:Model} we detail the neural network training workflow in PANNA; in Section~\ref{s:Implementation} we  provide computational details on the implementation, data format and parallelization strategies; in Section~\ref{s:Usage} we give a usage example; and finally in Section~\ref{s:Results} we report the results on two test cases to  show the capabilities of the package.

\begin{figure}[!hbt]
\begin{center}
 \includegraphics[width=1.0\textwidth]{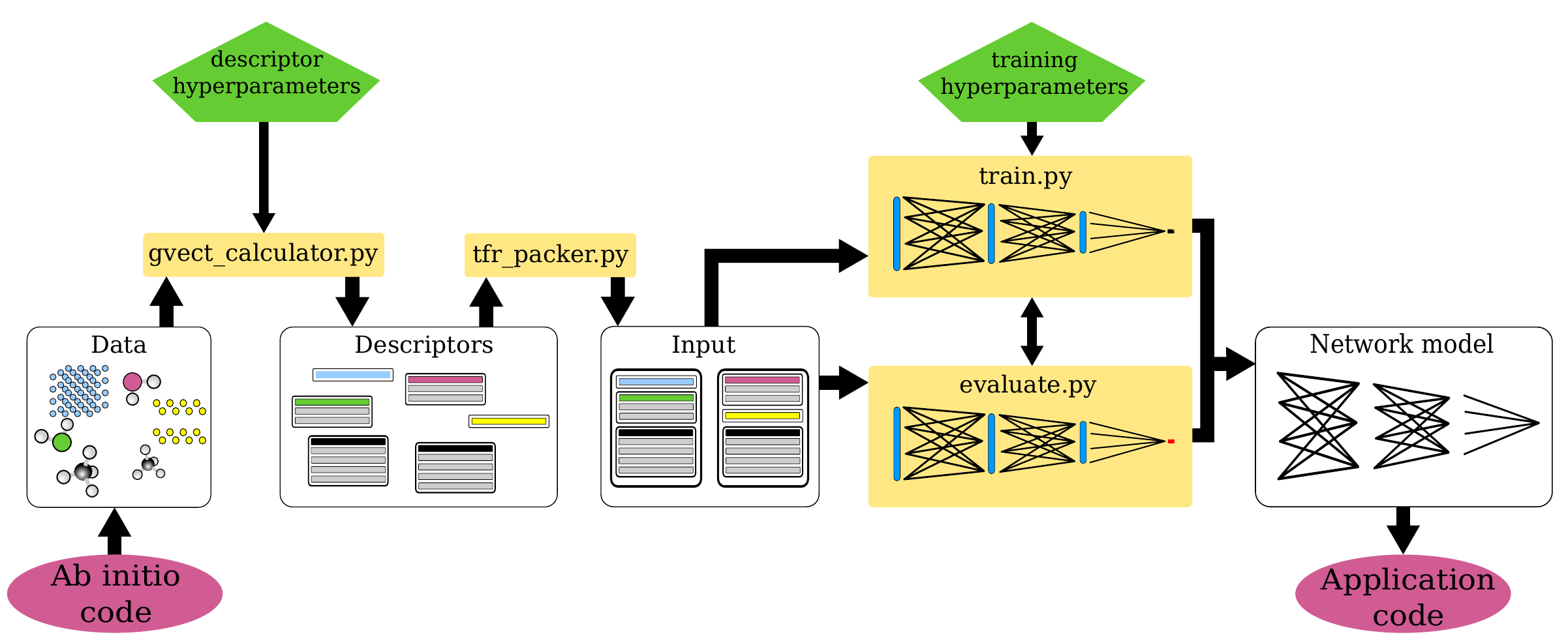}
\end{center}
\caption{PANNA workflow. Top layer corresponds to user intervention in the neural network potential generation process. Input and network hyperparameters are determined at this layer. Currently several helper programs (not shown) included in the PANNA package aim to help users in this intervention. In principle, the hyperparameters are also part of the network optimization process and can potentially become an automated part of the workflow. Yellow boxes indicate main programs in PANNA while white boxes stand for the user-owned data in different stages of processing and the network potential. The interface with the third party codes are established via parser scripts and patches (not shown) included in the package.}
\label{fig:PANNAWF}
\end{figure}


\section{Neural Network Training Workflow}\label{s:Model}

Here we summarize a simple neural network training workflow to introduce the notation used hereafter (see Fig.~\ref{fig:PANNAWF}). Consider a dataset of examples $D=\{s,\bar{O}\}$ where each example $s_i$ is associated with an observable of interest $\bar{O}_i$. For neural network interatomic potentials, these examples are often made up of atomic positions and the target observables are the corresponding total energies and atomic forces obtained from accurate first principles simulations.  
The first step of the workflow is to choose how to represent an example as input to the network model. We describe in section \ref{ss:Representation} the possible representations available in PANNA. Once a representation model is chosen, each simulation $s_i$ is mapped to an input vector $\vec{G}_i$ of length $N$ where $N$ is determined based on the description model and parametrization. The neural network can then be written as a parametrized function that maps input vectors into predictions for the target observable, $O_i=F(\vec {G}_i;W_{\rm NN})$, where $W_{\rm NN}$ stands for the parameter set. Finding the optimum set of parameters corresponds to approximating the computationally-demanding function that is used to generate the dataset $D$. The parameter fitting is performed via training of the network to minimize a cost function which is often chosen proportional to the difference between observable values in the data and the network predictions for them, $C(\{\bar{O}_i - O_i\})$.

Currently PANNA implements fully-connected, deep, feed-forward neural networks. All input vectors used in single training session are requested to be of same length, $N$. The network is composed of $L$ layers, each layer having $n_l$ nodes. Considering the input vector as the zeroth layer, we can define $n_0=N$. 
The computation performed at each node can be written in terms of its input, activation function, and output:
\begin{equation}\label{eq:activ}
    a_j^{l+1} = g\left(\sum_{i=1}^{n_l} a_i^l W_{ij}^{l+1} + b_j^{l+1}\right)
\end{equation}
where $a_i^{l}$ is the $i^{\rm th}$ element of the output from layer l, which is the input vector for node $j$ of layer $l+1$. The activation function for this node is specified by two parameters, weight matrix $W^{l+1}$ and bias scalar $b^{l+1}$, and a non-linear function $g$. 

The final output of the network $O_i = \vec{a}^L$ is therefore a complex nonlinear function of the input, weights and biases of each layer. If $g$ is partially differentiable with respect to weights and biases, these can be optimized to minimize the cost function $C$ via an optimization algorithm such as gradient descent (See section \ref{ss:Training}). Once the minimization is performed and a set of parameters that best estimate the observables for the training data --- as well as an independent validation data --- is found, the network can be used as a predictor for other examples (See section \ref{ss:Prediction}). 

In the following, we describe the choices readily available in PANNA for each step of this workflow.

\subsection{Representation}\label{ss:Representation}

Currently, two different types of representations are implemented in PANNA: the Behler-Parrinello (BP) descriptor model~\cite{BP07} and its recently modified version \cite{ANI17}. Both representations produce a fixed-size vector for each atom in the unit cell, describing its local environment up to a user-determined cutoff and with gaussian smoothing. 

The cutoff function ensures that for a given atom, only neighboring atoms closer than a cutoff radius $R_c$ have nonzero contribution to the descriptor vector and the contribution decays smoothly with distance. For a pair of atoms $i,j$ that are $R_{ij}$ apart, the cutoff function is: 
\begin{equation}\label{eq:fc}
    f_c(R_{ij})=\left\{
    \begin{array}{ll}
         \frac{1}{2}\left[
            \cos\left(\frac{\pi R_{ij}}{R_c}\right)+1
         \right]&R_{ij}\leq R_c  \\
         0& R_{ij}>R_c.
    \end{array}
    \right.
\end{equation}

In the standard BP representation, the descriptor vectors are made up of two parts: radial and angular. Radial part depends only on the interatomic distances from the central atom $i$ to all neighbours $j$ within the cutoff as:
\begin{equation}\label{eq:BPrad}
    G_i^{\rm Rad}[s]=\sum_{j\neq i}
    e^{-\eta\left(R_{ij}-R_s\right)^2}
    f_c(R_{ij}),
\end{equation}
where the width $\eta$ and a set of centers $R_s$ are parameters of the descriptor.
For the angular part, all pairs of neighbours $j, k$ in the cutoff radius that form an angle $\theta_{ijk}$ with the central atom $i$ is used to define:
\begin{align}
\label{eq:BPang}
G_i^{\rm Ang}[s]=&2^{1-\zeta_s}\sum_{j,k\neq i}
    \left(1+\lambda \cos \theta_{ijk} \right)^{\zeta_s} \nonumber \\
&\times e^{-\eta_s\left(R_{ij}^2+R_{ik}^2+R_{jk}^2\right)}\nonumber \\
&\times f_c(R_{ij})f_c(R_{ik})f_c(R_{jk}) 
\end{align}
where user-defined sets of $\eta_s$ and $\zeta_s$ are the parameters of the descriptor. The $\lambda$ parameter takes $\pm 1$ in order to generate descriptors that display peak response at $0$ and $\pi$.

In the modified BP representation, the angular descriptors are modified to contain the radial and angular resolution more explicitly:
\begin{align}\label{eq:mBPang}
G_i^{\rm Ang}[s]=&2^{1-\zeta}\sum_{j,k\neq i}
    \left[1+\cos\left(\theta_{ijk}-\theta_s\right)\right]^\zeta \nonumber\\
    &\times e^{-\eta\left[\frac{1}{2}\left(R_{ij}+R_{ik}\right)
    -R_s\right]^2} \nonumber \\
    &\times f_c(R_{ij})f_c(R_{ik})
\end{align}
with user-defined parameters $\eta$, $\zeta$, set of $\theta_s$ and set of $R_s$  as the parameters of the descriptor.

In case of a system with multiple atomic species, the descriptors are resolved by species, i.e. radial descriptor sum in Eq.~\ref{eq:BPrad} is performed only for neighbors belonging to a single species at a time and the total radial descriptor size grows linearly with number of species $n_s$. Similarly, the sum in angular descriptors are repeated for each possible pair of species for a given central atom, giving rise to a growth by $n_s(n_s+1)/2$ in size of the angular descriptor.

It should be noted that a good representation is fundamental for generating a successful neural network potential. For any representation model chosen, the descriptor parameters should be carefully selected in order to achieve an optimal trade off between richness of the descriptors and the accompanying computational cost during training. PANNA provides visual inspection tools to examine the resolution of the descriptors as well as the suitability of the descriptor parameters for a given dataset (See section~\ref{s:Implementation}).

\subsection{Network Architecture}\label{ss:Architecture}

PANNA constructs a sub-network for each atomic species such that the output of sub-networks can be combined to obtain a single prediction output for a given atomic configuration input. This is a well-established architecture and several works obtained with this architecture can be found in the literature \cite{Onat2018,ARTRITH2016135,LEE201995,KHORSHIDI2016310}. In order to address the varying complexity in the environment of different species in a sample, PANNA allows each sub-network to be of different size. Moreover, to enable tighter control of the training dynamics, the option to freeze any layer is also implemented, so that only user-selected parts of the network can be trained. Lastly, a different non-linear activation function $g$ can be chosen for each layer. Currently supported activation functions are:

\begin{itemize}
    \item Gaussian: $g(x)=\exp(-x^2)$, as demonstrated in Ref.~\cite{ANI17}.
    \item ReLU (rectified linear unit): $g(x)=\max(0,x)$.
    \item Linear: $g(x)=x$
    \item Radial Basis Function (RBF) that changes the structure of the layer from the one outlined in Eq.~\ref{eq:activ} into $a_j^{l+1} = \exp\left[-\sum_{i=1}^{n_l} (a_i^l-W_{ij}^{l+1})^2\right]$. 
\end{itemize}

The last layer of the network that results in the final estimate for the observable uses a linear activation in order to sum the partial predictions of each atomic species sub-network.

\subsection{Training}\label{ss:Training}
The training of the model parameters is performed through a commonly used variant of stochastic gradient descent called Adam~\cite{Adam} which uses the gradients of the cost function with respect to the weights to drive the model parameters towards a local minimum.
The stochasticity is a result of the practice that at each minimization step, only a randomly selected subset (or {\it minibatch}) of the whole training data are used for gradient calculation. The minibatch cost function $C_{\rm b}$ at optimization step $t$ is 
\begin{equation}
    C^{(t)}_{\rm b}(W^{(t)}) = \sum_{i\in {\rm batch^{(t)}}}C^{(t)}(\bar{O}_i-O_i(W^{(t)})).
\end{equation}
where the parameters of the model (weights and biases of all layers) at step $t$ is represented with $W^{(t)}$ for brevity.

Adam optimizer updates each parameter $w \in W$ adaptively, i.e. depending on its individual history until optimization step $t$. It first calculates an estimate for the mean $m_w$ and for the uncentered variance $v_w$ for the current and past gradients of each parameter via an exponential running average:
\begin{eqnarray}\label{eq:AdamMean}
m_w^{(t)} = (1-\beta_1)\nabla_wC^{(t)} + \beta_1 m_w^{(t-1)}\\
v_w^{(t)} = (1-\beta_2)|\nabla_wC^{(t)}|^2 + \beta_2 v_w^{(t-1)}
\end{eqnarray}
where $\beta_1, \beta_2 \in [0,1) $ are the exponential decay rates for the running average. The bigger they are, the more history dependent the optimization becomes. Adam also implements an effective correction in order to avoid the bias due to initial values where there is no history:
\begin{eqnarray}
\hat{m}_w^{(t)} = m_w^{(t)} / (1-(\beta_1)^t) \\
\hat{v}_w^{(t)} = v_w^{(t)} / (1-(\beta_2)^t)
\end{eqnarray}
Then it updates the parameters of the model proportionally with respect to the mean and inversely proportionally to the square root of the variance of their gradient so that parameters whose gradients show large variation in the last few steps are updated more slowly in the coming step and vice versa :
\begin{equation}\label{eq:AdamUpdate}
    w^{(t+1)} = w^{(t)} - \alpha \frac{\hat{m}^{(t)}_w}{\sqrt{\hat{v}^{(t)}_w}+\epsilon}
\end{equation}
where $\alpha$ is the learning rate and $\epsilon$ is a small number for stability. The most important parameters of the Adam optimizer for tuning of the training of a neural network model are $\alpha$, $\beta_1$ and $\beta_2$. Note that TensorFlow implementation of Adam differs slighly as in Eq.~\ref{eq:AdamUpdateTF}; resulting in scaling of $\epsilon$. For further details about the Adam algorithm see Ref.\cite{Adam}.

\begin{equation}\label{eq:AdamUpdateTF}
    w^{(t+1)} = w^{(t)} - \alpha \frac{(1-(\beta_2)^t)}{(1-(\beta_1)^t)} \frac{m^{(t)}_w}{\sqrt{v^{(t)}_w}+\epsilon'}.
\end{equation}

In addition to the adaptive stepsize nature of Adam, PANNA allows to gradually decrease the learning rate $\alpha$ during training to mimic annealing in the parameter space. The exponential decrease of the learning rate is given by: 
\begin{equation}\label{decay_lr}
    \alpha^{(t)} = \alpha^{(t=0)} r^{t/\tau},
\end{equation}
 where $\alpha^{(t)}$ is the decayed learning rate at step $t$. The decay rate $r$ and the decay step $\tau$ are used to determine the decay behavior.

The cost function is an important ingredient of neural network training. In PANNA besides the standard quadratic loss on target observable values (Eq.~\ref{eq:quadloss}), it is possible to use an exponential loss function, which weights more strongly the outliers within a batch.
To avoid numerical instability at the initial steps of the training when the gradients can be expected to be large, the exponential loss is smoothly clipped to a constant value through the application of a hyperbolic tangent (see Eq.~\ref{eq:exploss}). If desired, cost per atom can be considered rather than cost per data point.
\begin{equation}\label{eq:quadloss}
    C_{\rm b}^{\rm Q}(W) = \sum_{i\in {\rm batch}} \left(\bar{O}_i-O_i(W) \right)^2.
\end{equation}  
\begin{equation}\label{eq:exploss}
    C_{\rm b}^{\rm E}(W) = \exp\left[ a \tanh \left(\frac{1}{a} \sum_{i\in {\rm batch}}\left(\frac{\bar{O}_i-O_i(W)}{N^{\rm atoms}_i} \right)^2\right)\right ].
\end{equation}

Adding further constraint or information to cost function, i.e. regularization, can help to prevent overfitting. Two commonly used norm-based weight regularizations are supported in PANNA: regularization on the sum of the absolute value (L1 norm) and on the square (L2 norm) of the weights. The relative coefficient for each norm can be set independently. Resulting penalty is added to the total cost in the computation of gradients (Eq.~\ref{eq:normloss}).
\begin{equation}\label{eq:normloss}
    C_{\rm b}^{\rm NR}(W) =  
   c_1 \left \Vert W \right \Vert_1 + 
   c_2 \left \Vert W \right \Vert_2
\end{equation} 

On occasions where parameter space have steep cliffs, gradients can reach large numbers and following them may result in overshooting the low cost region. To prevent this behavior, the absolute value of each gradient can be capped at a user-defined constant before being processed to update the parameters. It should be noted that such gradient clipping based on absolute value corresponds to change of optimization direction in the parameter space. 

\subsection{Training with Forces}\label{ss:TrainingForces}

When both the target function and its derivatives are known and used in the training, the prediction power of a neural network model can dramatically increase. Since training already requires differentiation with respect to network parameters, additional cost of training for the derivatives can be mitigated with the chain rule. In atomic simulations, this scenario can be realized when the cost function includes both energy and forces as their analytical derivative with respect to atomic positions. 

In PANNA, force prediction of the network model is computed analytically as in the following:
\begin{equation}
    \vec{F}_i = \sum_{jk} \frac{\partial E_j}{\partial G_{jk}}
     \frac{\partial G_{jk}}{\partial \vec{x}_i},
\end{equation}
where $G_{jk}$ is the $k$-th element of the descriptor for atom $j$. The first term is already required during training for the update of network parameters (see Eq.~\ref{eq:AdamMean}), while the second term of partial derivatives with respect to the position of atom $i$ can be pre-computed together with the descriptors (see section~\ref{ss:Representation}).

In PANNA, when reference forces are requested to be used in training, an extra term is added to the cost function:
\begin{equation}\label{eq:Floss}
    C_{\rm b}^F = c_F \sum_{i \in \rm{batch}}\sum_{j=1}^{N_{\rm atoms}} 
    \sum_{k\in \{x,y,z\}} (\bar{F}_{ij}^k - F_{ij}^k)^2,
\end{equation}
where $c_F$ is a user-defined parameter that allows to tune the weight of the force-based loss in the total cost function.

\subsection{Prediction}\label{ss:Prediction}
Once the network is trained, prediction on a new data point can be computed through a single forward evaluation of the network.
Unlike training, where powerful libraries are necessary for efficient optimization of parameters and data-handling of large sets of data, the evaluation task only requires matrix multiplication and basic mathematical algebra. This enables PANNA to provide a simple program for the evaluation task, which can be easily coupled with other codes, e.g. MD software that simulates movement of the atoms as directed by the interatomic neural network potential. Currently PANNA supports such interface directly with LAMMPS code \cite{LAMMPS,footnote1} and indirectly through KIM API ~\cite{KIM1,KIM2,panna-kim}.

\section{Implementation details}\label{s:Implementation}
\subsection{PANNA Programs}\label{ss:PANNAPrograms}
PANNA core is written in Python and is based on the TensorFlow (TF)~\cite{TF} framework. It is organized as a package of several main programs, each characterized by its own configuration input file (.ini) and parameters. In section \ref{s:Usage} we offer a usage and input example for each of these programs:
\begin{itemize}
    \item \begin{verbatim}gvect_calculator.py\end{verbatim}
    This code first reads information on the simulation data such as atomic positions and cell parameters, written in simple PANNA-example format, stored as JSON standard. It then computes the descriptor for each atom, with the user-specified hyperparameters, and stores all the descriptors of all atoms in the simulation in a single binary data file.
    
    \item \begin{verbatim}tfr_packer.py\end{verbatim} 
    This code collects a large number of descriptor binary files and converts them into TFRecord (TFR) format which is ready to be efficiently processed by TF. The resulting files, called TFData within PANNA, reduce the I/O overhead, simplify and speed up the dataset management during training. 
    
    \item \begin{verbatim}train.py\end{verbatim}
    This is the main routine that performs the training: it reads the TFData files, handles the queue management to supply parallel processing of minibatches and drives the training procedure with the appropriate TF calls. Information required to restart the calculation is stored as ``checkpoints'' during training in TF format at user-defined intervals. Additionally, the summary of each training is stored in TF ``event'' files that can be visualized in TensorBoard~\cite{TF} (see section~\ref{ss:Visualization}).
    
    \item \begin{verbatim}evaluate.py\end{verbatim}
    Predictions are made using this code, which can parse the checkpoints, access the parameters of the network and evaluate the network prediction for a given input configuration. It can operate on single binary descriptor files, such as the outputs of \texttt{gvect\_calculator.py} or bundled TFData files such as the outputs of \texttt{tfr\_packer.py}. It can operate on a single file or on all files of a directory as specified by user input.
\end{itemize}

A number of tools are also provided to handle data processing and visualization. Just to name a few: parsing of Quantum ESPRESSO output XML files~\cite{QE17}, of extended xyz format as in Ref.\cite{Wen-KLIFF}, or of ANI~\cite{ANI17} dataset are done by \texttt{qe\_parser.py}, \texttt{exyz\_parser.py}, \texttt{ani\_parser.py} respectively. These tools convert the data into PANNA-example files in JSON standard. Multiple such example files can be converted into an XCrySDen~\cite{KOKALJ1999176} animation by \texttt{json2axsf.py} for visual inspection of the data. The \texttt{gvect\_param\_check.py} code plots the characteristic length scales of a descriptor to help tune the radial and angular resolution parameters. Given a set of PANNA-example files and descriptor hyperparameters, \texttt{gvect\_writer.py} calculates the average descriptor for each species, which may prove useful in recognizing the optimum parameter range for a specific set of data. Once the training is over and a network model is decided to be used in further applications, \texttt{extract\_weights.py} program converts the network architecture and weights from tensors in TF to PANNA, LAMMPS or KIM compatible format.

\subsection{TensorFlow}\label{ss:Tensorflow}
TensorFlow (TF), the underlying engine for data management and neural network training used in PANNA, is an open source machine learning library released by Google~\cite{TF}. 
In TF, a computation is described by a directed graph that represents values flowing between nodes. Each node represents an instantiation of an operation, such as matrix multiplication, and has zero or more inputs and outputs. Some nodes are allowed to maintain and update their persistent states, to enable branching or looping as needed to represent the computation. In dataflow programming, movement of data is emphasized: a node operation runs as soon as the inputs become valid, making this programming paradigm suitable for distributed parallel execution. In TF all data are treated as n-dimensional arrays, i.e. tensors.

Due to its dataflow programming basis, TF is particularly well optimized for management of large datasets and training of large networks. Its stateful queue operations support advanced form of coordination of access to data. It allows a different access for pre-fetching and pre-processing of training examples, for shuffling them, and for consuming them during training, allowing user to fine tune the dataflow in their application. Several such performance related TF variables are exposed in PANNA through input keywords such as \texttt{shuffle\_buffer\_size\_multiplier} which sets how many minibatches of data are to be accessed for shuffling. Another of such examples is the \texttt{dataset\_cache} boolean that is used to determine whether to cache the TF input dataset during training for increased performance, noting that TF dataset is a complex collection of elements including the data but also possible operation objects that act on the data such as iterators.

\subsection{Parallelization and Hardware}\label{ss:Parallelization}

TensorFlow and PANNA performance can be controlled by two main variables: \texttt{intra\_op\_parallelism\_threads} and \texttt{inter\_op\_parallelism\_threads}. The \texttt{intra\_op} keyword sets the parallelism inside an operation, e.g. when a matrix multiplication is performed; while \texttt{inter\_op} controls the parallelism of operations that can be carried out concurrently. PANNA exposes these variables with the same input keywords as TF. While the optimum number of threads for intra operation parallelization can be expected to be in the same order as other codes that perform matrix manipulation --- and users can benefit from benchmarks based on \texttt{\$OMP\_NUM\_THREAD} as the parallelization variable --- the inter operation parallelism is largely affected by the network size and topology, so that its optimization per training scenario is advised.

Thanks to the underlying TF engine, PANNA can run on CPU, GPU and TPU systems. An example is provided in the documentation to demonstrate how to run PANNA with intra- and inter-node parallelization on multiple CPU nodes of a High Performance Computing (HPC) system.

Each operation of TF resides on a particular device and task, e.g. \texttt{Stitch} operation that reassambles partial results may be required for a \textit{Parameter Server} task that contains one CPU device, or \texttt{Add} operation for tensors of a \textit{Worker} task that contains two devices, one of which is the same of the parameter server task before, and one GPU device. TF first places the operations in a graph to devices, then partitions all operations of a device into subgraphs that can be cached, and manages communication between devices via \texttt{Send} and \texttt{Recv} operations that are specialized and optimized for several device-type pairs. 
PANNA modifies the device-based parallelization of TF such that every worker task executes the same neural network training graph independently. During training, workers read the network parameters and copy them internally, process a minibatch of data and calculate the gradients required for optimization, then send the gradients to the parameter server that updates the single shared instance of the network parameters independently. Therefore the weights and biases of the network are updated asynchronously, reading and updating of different workers happen concurrently, making asynchronous training free from read-write lock, enabling high scalability across nodes. It should be noted that asynchronous update introduces further stochasticity to the training procedure since at any moment network parameters are allowed to have different time stamps \cite{NIPS2012_4687}.  

\subsection{Visualization}\label{ss:Visualization}
As the NN training involves a stochastic process with many parameters and figures of merit, tools that are easy to use for monitoring training dynamics are highly desirable.

\begin{figure}[!h]
\begin{center}
 \includegraphics[width=0.5\textwidth]{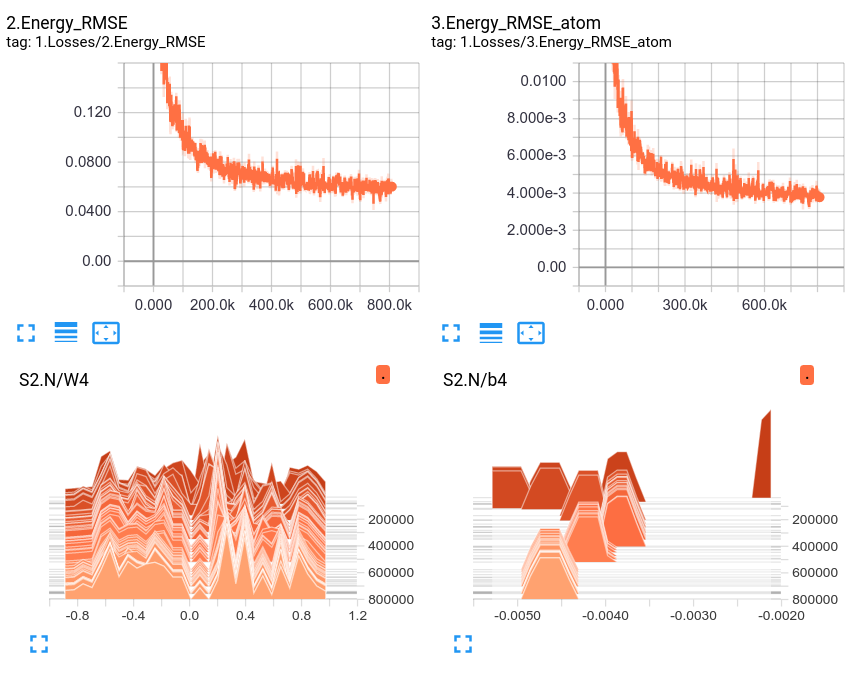}
\end{center}
\caption{Example of visualization in TensorBoard. {\bf Top:} Scalars such as loss per atom, regularization loss, or learning rate can be tracked as the training continues. {\bf Bottom:} Distribution of weights and biases of each layer and each atomic species as a function of optimization step can be observed. Such observations can give insight on when to try training strategies such as (un)freezing layers. }
\label{fig:TBEx}
\end{figure}

TensorBoard (TB), a TF-native visualization tool, accesses log files created at training time and visualizes them through a browser interface. In PANNA, a number of quantities that are relevant for NN training are saved in the log files such as total/per atom/root-mean-square of cost function per minibatch, loss due to L1 and L2 penalties, species-resolved atomic contribution to network prediction, the distribution of weights and biases for each layer, and the learning rate. See Fig.~\ref{fig:TBEx} for a sample view of elements from TB dashboard during training.

TB also includes dimensionality reduction tools such as Principal Component Analysis~\cite{PCA} and t-Stochastic Neighbor Embedding~\cite{tSNE} on the network parameters and allows the results to be visually analyzed.

\section{Usage example}\label{s:Usage}
In this section we describe a simple but complete use case of PANNA, from the atomistic simulation information to obtaining an NN interatomic potential ready to be employed in MD simulations. Simple input scripts are provided. Further details can be found in the tutorials distributed within the package.
\subsection{Data preparation}
Let us assume to have ab initio simulation results for water molecule obtained by Quantum ESPRESSO~\cite{QE17}. This package outputs in XML format. A relevant snippet of the XML file is as follows:


\begin{verbatim}
{
 <control_variables>
      <prefix>H2O_3e45a0e332</prefix>
      ...
 </control_variables>
 ...
 <atomic_structure nat="3" alat="3.779452265771e1" 
 bravais_index="1">
      <atomic_positions>
        <atom name="O" index="1">-3.103910095932729e-18
        1.972216586331875e-3 1.829760905993973e-1</atom>
        <atom name="H" index="2">3.739048037144415e-17
        1.613226163461667e0 -5.580297760571020e-1</atom>
        <atom name="H" index="3">1.187082814066741e-17
        -1.644526707907135e0 -5.824461493348307e-1</atom>
      </atomic_positions>
      <cell>
        <a1>3.779452265771287e1 0.000000000000000e0
        0.000000000000000e0</a1>
        <a2>0.000000000000000e0 3.779452265771287e1
        0.000000000000000e0</a2>
        <a3>0.000000000000000e0 0.000000000000000e0
        3.779452265771287e1</a3>
      </cell>
    </atomic_structure>
    ...
    <total_energy>
      <etot>-2.200687664809398e1</etot>
      ...
    </total_energy>
}
\end{verbatim}

Using \texttt{qe\_parser.py} , the relevant information, i.e. atomic positions, cell parameters and total energy, is parsed and stored in a JSON standard for neural network training:

\begin{verbatim}python3 panna/tools/qe_parser.py 
                 -i /path/to/QE_XML_FILES 
                 -o /path/to/PANNA_JSON_FILES 
                 --addhash \end{verbatim}
where input and output directories are used to process multiple files at once. The resulting JSON file is as follows:

\begin{verbatim}
{
  "atomic_position_unit": "cartesian", 
  "lattice_vectors": 
    [[37.79452265771287, 0.0, 0.0], 
    [0.0, 37.79452265771287, 0.0], 
    [0.0, 0.0, 37.79452265771287]], 
  "energy": [-22.00687664809398, "Ha"],
  "atoms": 
    [[1, "O", 
      [-3.103910095932729e-18, 0.001972216586331875,
      0.1829760905993973], 
      [-1.811290045808234e-06, -0.02306295358423087,
      0.05599605952219937]], 
    [2, "H", 
      [3.739048037144415e-17, 1.613226163461667,
      -0.558029776057102],
      [9.212452357485834e-07, 0.008691667402484345,
      -0.03363661323243399]], 
    [3, "H", 
      [1.187082814066741e-17, -1.644526707907135,
      -0.5824461493348307], 
      [8.900448100596506e-07, 0.01437128618174652,
      -0.02235944628976539]]], 
  "key": "H2O_3e45a0e332", 
  "unit_of_length": "bohr"
}
\end{verbatim}
When processing many \textit{ab initio} simulation at once, JSON files should be named carefully to avoid overwriting. The \texttt{--addhash} argument ensures that even when ab initio filenames are the same, e.g. the QE default \texttt{prefix.xml}, JSON files are uniquely named using a hash function. Without this argument, the names of the XML files are kept for JSON files. 

\subsection{Computation of descriptors}\label{ss:U:gvect}
Computation of the descriptors for each example can be achieved as the following:
\begin{verbatim}python3 panna/gvect_calculator.py --config gvect_input.ini\end{verbatim}
where a sample configuration file might look like the following:
\begin{verbatim}
[IO_INFORMATION]
input_json_dir = ./simulations
output_gvect_dir = ./gvectors
log_dir = .

[SYMMETRY_FUNCTION]
type = mBP
species = H, O

[PARALLELIZATION]
number_of_process = 4

[GVECT_PARAMETERS]
gvect_parameters_unit = angstrom
eta_rad = 16
Rc_rad = 4.6 
Rs0_rad = 0.5
RsN_rad = 16
eta_ang = 6.0
zeta = 50.0
Rc_ang = 3.1
Rs0_ang = 0.5
RsN_ang = 4
ThetasN = 8
\end{verbatim}
Here, the first section indicates the location of input and output directories, the second specifies the type of descriptors and atomic species, the third how many parallel processes to use and the fourth the hyperparameters used to create the descriptors.

In this case, modified BP descriptors will be created according to Eq. ~\ref{eq:BPrad} and Eq.~\ref{eq:mBPang}. The radial descriptor is made up of  \texttt{RsN\_rad} Gaussian centers equally spaced between \texttt{Rs0\_rad} and \texttt{Rc\_rad}. A similar relationship holds for the radial binning of the angular descriptor. For the angular binning, the centers are equally spaced in [0, $\pi$] window, shifted to align with the midpoint of the interval. Considering the $2$ species, this sample setting amounts to $16\times2$ radial components and $(4\times8)\times 3$ angular ones, i.e. a descriptor array of length $128$ for each atom.

The descriptor of each atom of the system is then concatenated to produce the descriptor of the simulation and written in binary format alongside the reference energy.

\subsection{Packing of data}\label{ss:U:tfr}
The large number of binary files produced in the previous step is packed into a small number of large files to increase I/O efficiency. This is achieved by the following command which yields binary files with .tfr extension.

\begin{verbatim}python3 panna/tfr_packer.py --config tfr_sample.ini\end{verbatim}
with a sample input as in the following:
\begin{verbatim}
[IO_INFORMATION]
input_dir = ./gvectors
output_dir = ./tfr
elements_per_file = 1000
prefix = train

[CONTENT_INFORMATION]
n_species = 2
\end{verbatim}

\subsection{Training}\label{ss:U:train}
A common practice is to divide the data into training and validation sets. Since in the previous step, we have packed all the data into self-contained .tfr files, this division can be performed simply by moving files into different directories. 
When the training set is created as such, the training process can begin. This can be achieved by the following program:
\begin{verbatim}python3 panna/train.py --config train.ini \end{verbatim}
A minimum required input file for this program is as follows:
\begin{verbatim}
[IO_INFORMATION]
data_dir = ./tfr_train
train_dir = ./train
log_frequency = 10
save_checkpoint_steps = 500

[DATA_INFORMATION]
atomic_sequence = H, O
output_offset = -13.62, -2041.84

[TRAINING_PARAMETERS]
batch_size = 50
learning_rate = 0.01
max_steps = 5000

[DEFAULT_NETWORK]
g_size = 128
architecture = 64:32:1
trainable = 1:1:1

[H]
architecture = 32:1
trainable = 1:1
activations = ReLU:Linear
\end{verbatim}

In the first section the necessary information on I/O paths and logging frequency is specified. In the second, the atomic species sequence used to build the descriptors as well as a reference energy for each species is given. The reference energy listed here will be subtracted from the total energy of any data point before the data are presented to the neural network. Even though this operation only corresponds to a trivial shift of bias, hence the network can be expected to learn its value during training; because the values of the network parameters are initialized around zero, explicitly applying the shift is found to speed up the training process. The third section contains information about the specifics of the training: the size of the minibatch, the learning rate to use (constant in this case) and how many training steps to perform before stopping. 

In the following section a default network model is defined. This is the model that would be assumed by default for each species unless further information is provided. The size of the input descriptor \texttt{g\_size}=128 and the architecture (e.g., two hidden and one final output layer) are specified. The activation function is by default Gaussian for all the hidden layers, and all layers will be allowed to change during training. Finally the last section allows the user to define a different network model for a species of his/her choosing.
This feature can be particularly useful for fine tuning the network in order to achieve a compact, low cost model for several species. 

At any time during or after the training, we could inspect the evolution of the observables by using TB:
\begin{verbatim}tensorboard --logdir=./train \end{verbatim}

\subsection{Validation of the model}\label{ss:U:validate}
The performance of a neural network model in terms of accuracy can be estimated through the evaluation of the final model on data that has not been included in the training set.
This can be achieved with the following program:
\begin{verbatim}python3 panna/evaluate.py --config validation.ini \end{verbatim}
and sample input file:
\begin{verbatim}
[IO_INFORMATION]
data_dir = ./tfr_validate
train_dir = ./train
eval_dir = ./validate

[DATA_INFORMATION]
g_size = 128
atomic_sequence = H, O
output_offset = -13.62, -2041.84

[VALIDATION_OPTIONS]
single_step = True
\end{verbatim}
The locations of the validation data, train files, i.e. weights and biases, and validation output are given in the first section. The second section specifies the network-related properties of the validation data, which is expected to be coherent with the specifications used during training. The exception to this is \texttt{output\_offset}, e.g. the validation data might be resulting from a total energy calculation with a different zero for energy, for example, due to a different implementation, pseudopotential or exchange correlation functional being used.

This program produces a simple text file with reference and predicted output for all the simulations in the validation dataset, and can be used to asses the quality of the model.

\section{Results}\label{s:Results}

In this section we detail two example studies that demonstrate the usability of PANNA for periodic and aperiodic systems, with varying amount and quality of data. Finally we also demonstrate the energy conservation of the final network model in an MD scenario. 

\subsection{Molecules}\label{ss:Molecules}
Here we report training of the network architecture previously used in Ref.~\cite{ANI17}. With half a million parameters, this is a larger network than the average architecture employed in the literature. By reproducing the results of Ref.~\cite{ANI17}, we demonstrate that the training and testing modules of PANNA can answer the high performance demand scenarios in machine learning of interatomic potentials. 

The dataset~\cite{Smith2017-ANIdata} contains $57462$ small organic molecules consisting of H, C, N and O atoms, with up to 8 heavy atoms and corresponding total energy calculated via Density Functional Theory (DFT) \cite{DFT-HK}. Training is performed for 3 different datasets, labeled as DSmax4, DSmax6 and DSmax8, each including data from molecules with up to 4, 6, or 8 heavy atoms respectively. Table~\ref{table:Ani_train_valid_sizes} shows the size of datasets used for training and validation.
\begin{table}[!hbt]
    \centering
    \begin{tabular}{|c|c|c|c|}
    \hline
    Label &
     \begin{tabular}[x]{c}
          Max \# of \\
          heavy \\
          atoms
     \end{tabular} & \begin{tabular}[x]{c}
     \# of elements \\
     in training \\
     set $[\times10^6]$
     \end{tabular} & 
     \begin{tabular}{c}
           \# of elements \\
          in validation  \\
          set $[\times10^6]$
     \end{tabular}\\
     \hline
     DSmax4 & 4 &0.656&0.134\\
     DSmax6 & 6&3.432&0.427\\
     DSmax8 & 8&17.476&2.182\\
     \hline
    \end{tabular}
    \caption{The size of datasets obtained from Ref.\cite{Smith2017-ANIdata} used for training and validation. As the dataset is constructed by  sampling the normal modes of each molecule, alongside a scaling factor  to reduce the bias towards bigger molecules, it contains a different amount of data for each molecule type, e.g. 480 examples for $N_2$ and 17280 for $C_4H_{10}$ (butane) and 340 for $C_8H_{18}$(octane). The final models are benchmarked against 10347 configurations from normal mode sampling of 138 molecules from the GDB-11~\cite{GDB11} with 10 heavy atoms, also included in Ref.~\cite{Smith2017-ANIdata}. }
    \label{table:Ani_train_valid_sizes}
\end{table}

The modified Behler-Parrinello symmetry functions described in Section~\ref{ss:Representation} are used as in the original reference, with 32 Gaussian centers for the radial part, and 8 angular and 8 radial centers for the angular part. Considering the 4 species in the dataset, this results in descriptors of size $768$ for each atom. The atomic network architecture consists of 3 hidden layers of sizes $128$, $128$ and $64$, all with Gaussian activation function, followed by a linear activation layer.

While the originally proposed cost function is proportional to the exponential of the square loss, it is found to yield very large gradients in the early stages of the training causing numerical instability. The original reference addresses this by weights norm clipping. 
In this study a simple quadratic loss in combination with the capped exponential loss described in Eq.~\ref{eq:exploss} is found to alleviate the problem while still preserving increased gradients on the outliers. An initial learning rate of 0.001 with a decay rate $r=0.98$ and decay step $\tau=3200$ is used following Eq.~\ref{decay_lr}. In the case of training with DSmax8, learning rate decay step is increased to $\tau=16000$, leading to a slower decay. A fixed batch size of $1024$ examples is used.

\begin{figure}
    \centering
    \includegraphics[width=0.5\textwidth]{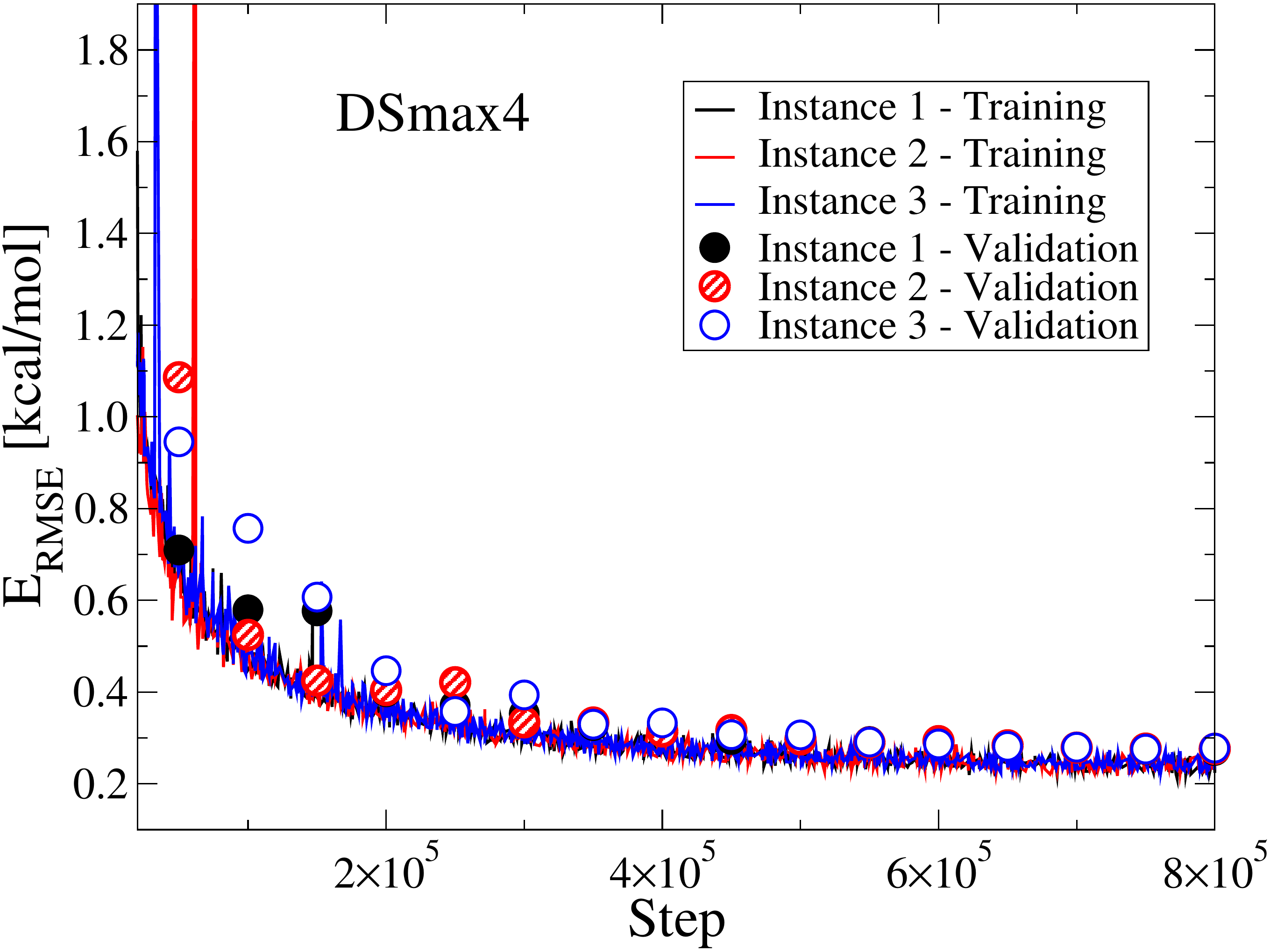}
    \includegraphics[width=0.5\textwidth]{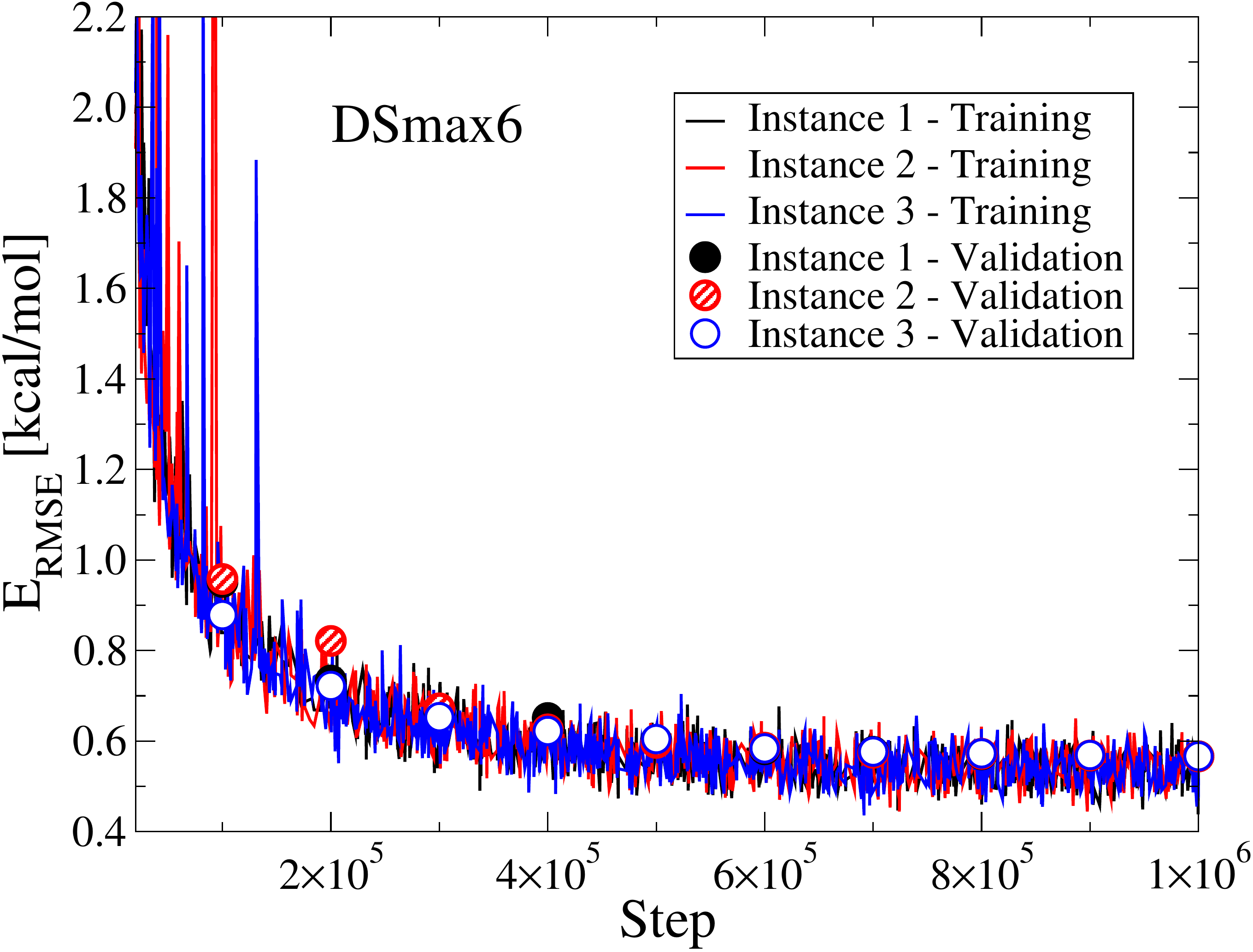}
    \includegraphics[width=0.5\textwidth]{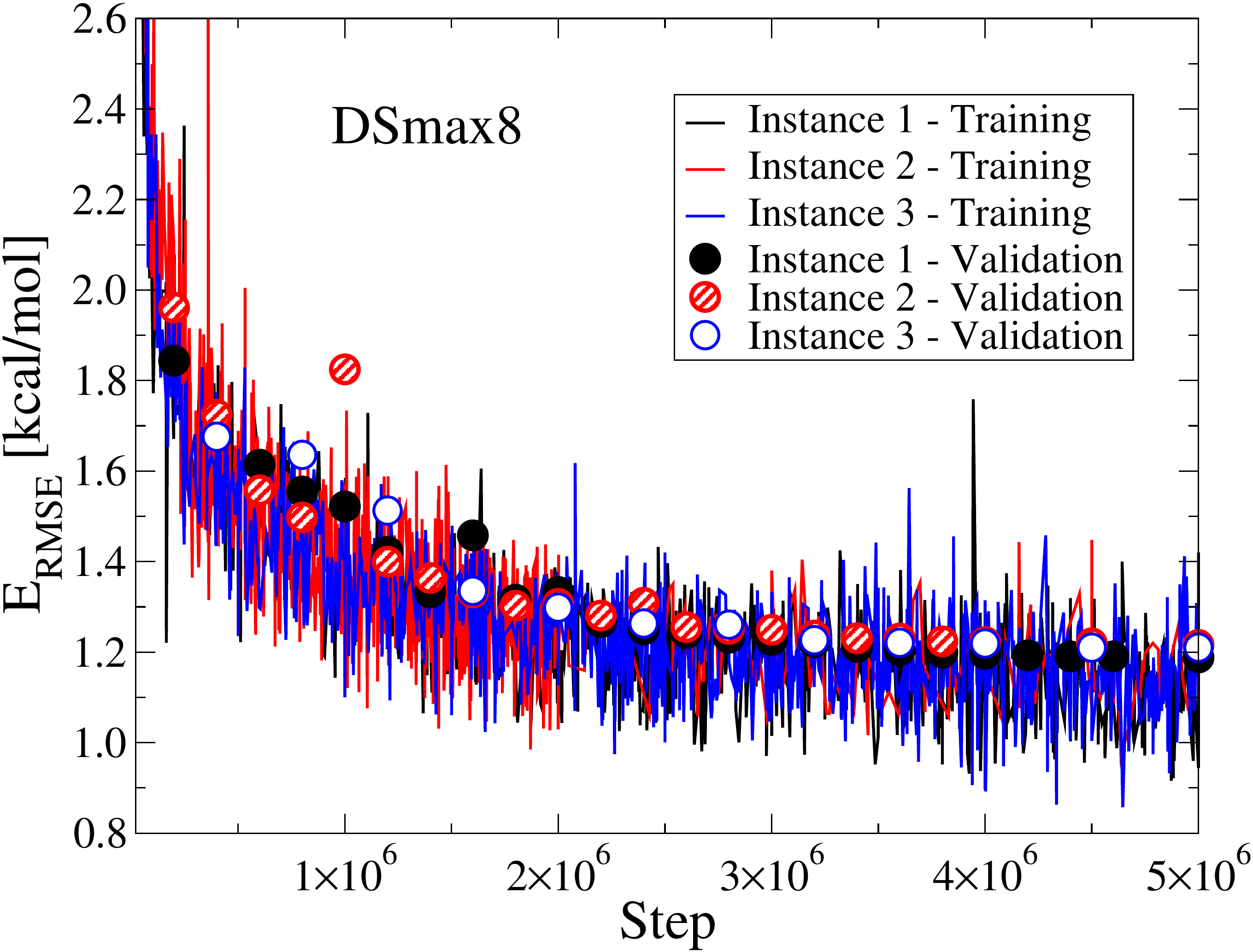}
    \caption{The energy RMSE during training and on validation set as a function of optimization steps for trainings with DSmax4 (top), DSmax6 (middle) and DSmax8 datasets (bottom). For each dataset, three instances of training is performed starting from different random initial parameters. 
    The RMSE calculated at the final step for training (validation) set are 0.24 (0.27), 0.24 (0.28) and 0.24 (0.28) ;  0.54 (0.57), 0.54 (0.57) and 0.53 (0.57); and 1.12 (1.19), 1.18 (1.22) and 1.14 (1.21) in kcal/mol for trainings with DSmax4, DSmax6 and DSmax8 respectively. For comparison, the results from Ref.\cite{ANI17} are 1.16 (1.28)~kcal/mol training (validation) RMSE for training with DSmax8.
    }
    \label{fig:468haplot}
\end{figure}

Figure \ref{fig:468haplot} shows the evolution of root mean square error (RMSE) on training and validation sets during training. For each dataset, three networks with identical architecture but different random seeds are trained. It can be seen that the proposed training schedule yields quantitatively reproducible results, which are consistently in good agreement with those reported in Ref.~\cite{ANI17}. 

As it may be expected, the bigger the training dataset gets in variability, going from DSmax4 to DSmax8, the harder it gets to solve the regression problem successfully in few optimization steps. Hence the training and validation errors increase steadily from approximately 0.3 to 1.2 kcal/mol per example, while the optimization steps required goes from approximately one to five million steps. 

The arithmetic average of different instances of trainings can be used to make committee predictions for each data point. The validation set RMSE resulting from such committees are 
0.25, 0.52, 1.14 for DSmax4, DSmax6, DSmax8 respectively. Note that these values are very close to best individual network prediction errors 0.27, 0.57, 1.19 respectively (See also Fig.~\ref{fig:468haplot}), hinting that energy prediction error of different networks for each example may be highly correlated. Single network predictions will be reported in the rest of this section.

While the above analysis is based on training and validation sets of similar complexity, the value of neural network potentials can be better judged based on their transferability.
To assess this property, we test networks on test sets of varying complexity. For example, a network trained on DSmax4 is tested on molecules with 5, 6, 7, 8 and 10 heavy atoms. In order to compare our results with Ref.~\cite{ANI17} in the case of 10 heavy atom test set, the RMSE is also calculated with respect to the lowest energy structure for each molecule. RMSE calculated this way is referred as relative RMSE. The summary of results is reported in Table~\ref{table:GDB_test}. It can be seen from the energy-capped vs uncapped results that transferability is increased when considering only the low energy structures, and the overall network performance reduces as the training and test sets become more dissimilar.

\begin{table}[!hbt]
    \centering
    \begin{tabular}{|c||*{8}{c|}}
    \hline
    \backslashbox{Training \\Set}{Test Set}
    &\makebox[1.5em]{DS5}
    &\makebox[1.5em]{DS6}
    &\makebox[1.5em]{DS7}
    &\makebox[1.5em]{DS8}
    &\makebox[1.5em]{DS10}
    &\makebox[1.5em]{\begin{tabular}{c}DS10\\all\end{tabular}}
    &\makebox[1.5em]{\begin{tabular}{c}DS10\\300\end{tabular}}
    &\makebox[1.5em]{\begin{tabular}{c}Ref.\\\cite{ANI17}\end{tabular}}\\
    \hline
    DSmax4 & 17.1&  23.2&28.3&30.0&24.5& 139.2 & 21.0&26.0 \\
    DSmax6 & 0.5& 0.7&13.5&15.5&14.5&138.5 &15.5 &17.7\\
    DSmax8 & 0.6& 0.7&1.2&1.4&2.1& 87.7 & 2.0&1.8\\
    \hline
    \end{tabular}
    \caption{RMSE of energy prediction in kcal/mol for networks trained and tested with datasets of various molecular complexity. 
    In the first five columns the RMSE is calculated for configurations where the total energy is within $E_{\rm cut}=275$\,kcal/mol of the lowest energy configuration for each molecule. \textit{DS$\langle N \rangle$} stands for dataset with molecules having \textit{$\langle N\rangle $} heavy atoms.
    Additionally for the 10 heavy atom set, the RMSE for all configurations independent of their energies, and the relative RMSE for configurations within the lowest $E_{\rm cut}=300$\,kcal/mol window is also given respectively. The last two columns contain the directly comparable values.}
    \label{table:GDB_test}
\end{table}

\begin{figure}
    \centering
    \includegraphics[width=0.5\textwidth]{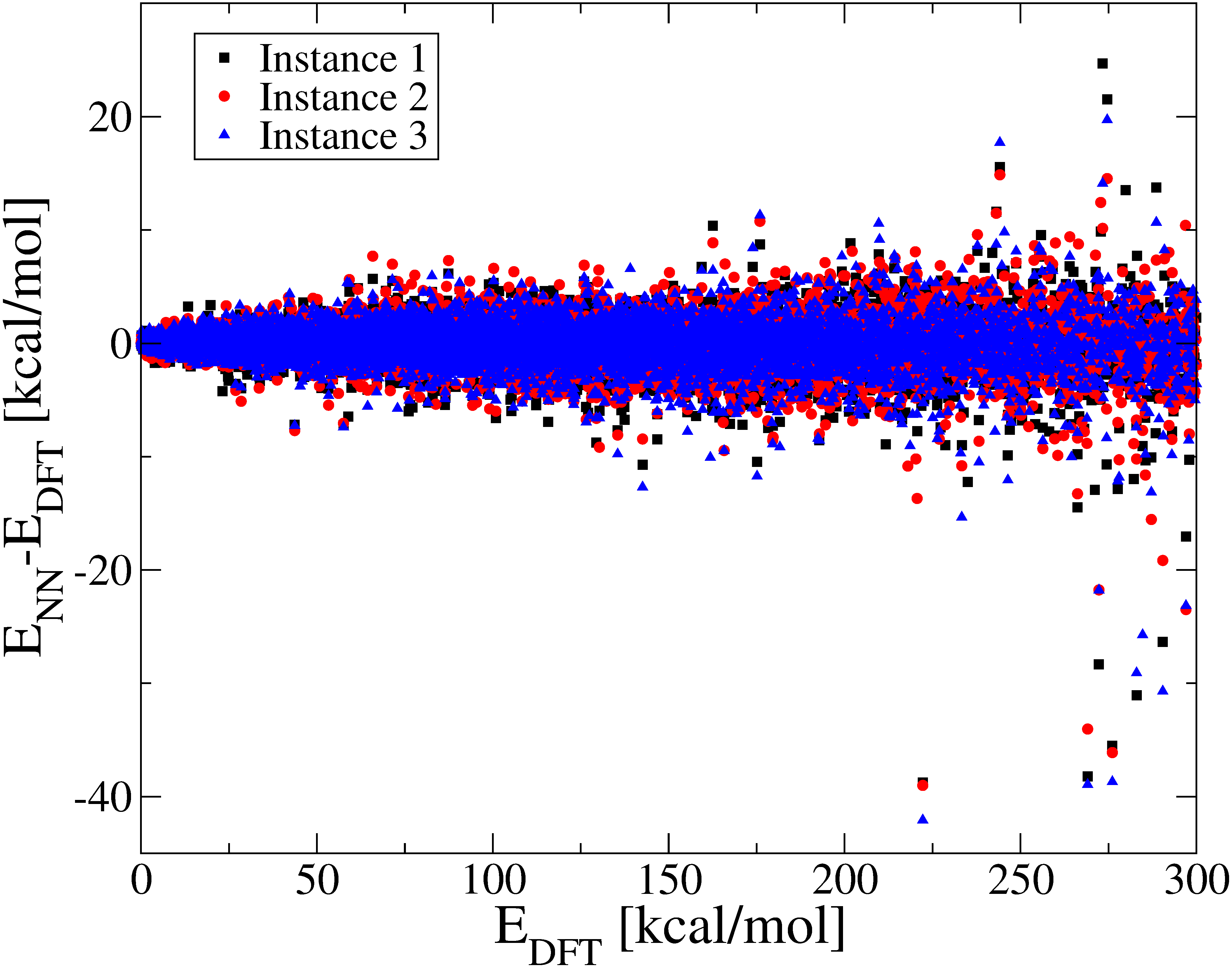}
    \includegraphics[width=0.5\textwidth]{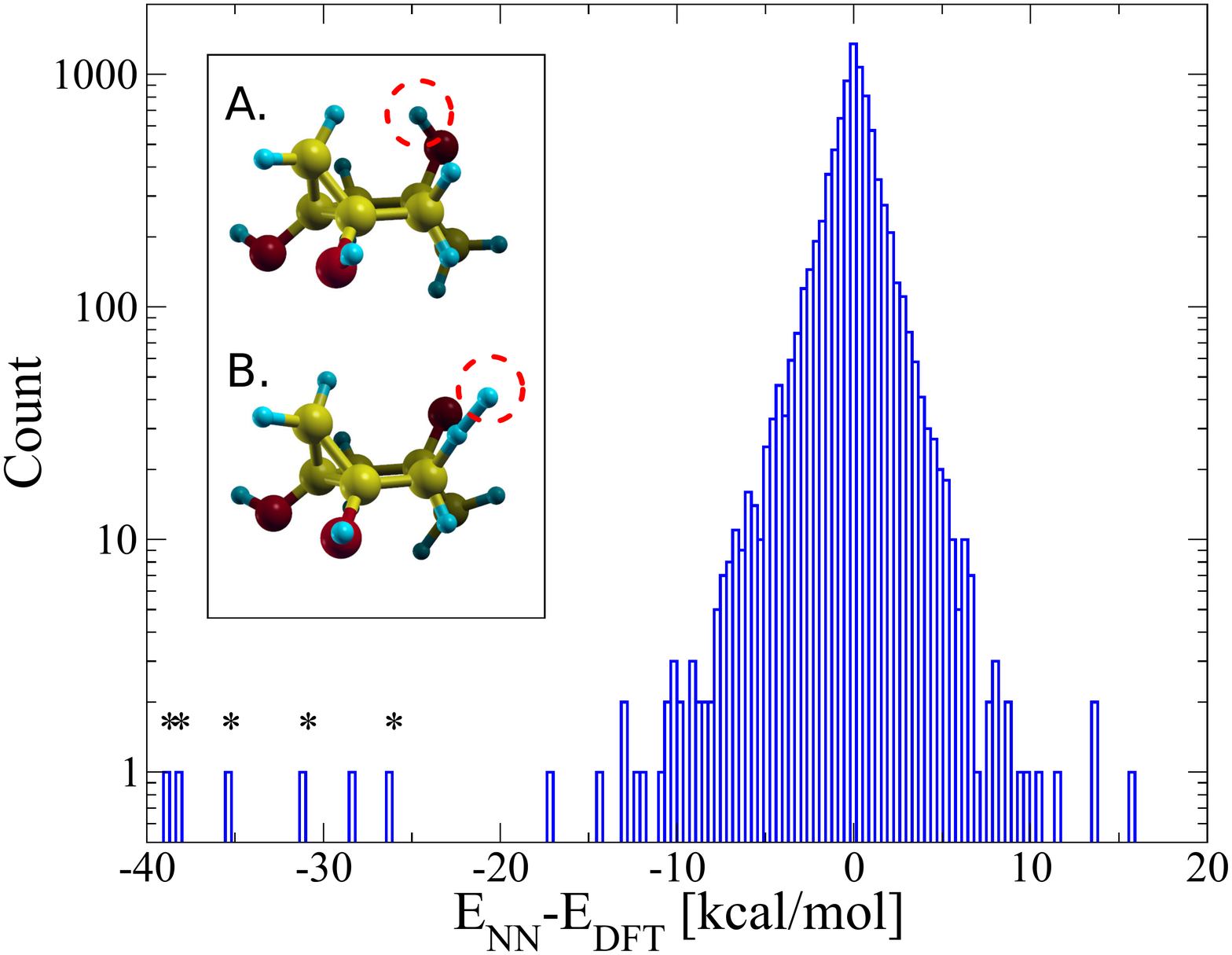}
    \caption{{\bf Top:} Neural network prediction error compared to DFT energy on GDB-11 set with 10 heavy atoms where total energy is shifted so that the lowest energy structure of each molecule in the dataset corresponds to 0~kcal/mol. The networks are trained on DSmax8 set (see Fig.~\ref{fig:468haplot}, bottom panel). 
    {\bf Bottom:} Prediction error distribution in log scale for Instance 2. It is noteworthy that all the marked outliers correspond to configurations of a single molecule shown in the inset where the H atom bound to O in the ground state configuration (A) is displaced far away, and may even form a bond with another H atom (B). Without the marked outliers, the RMSE reduces to 1.8 kcal/mol from 2.0 kcal/mol.} 
    \label{fig:ANI8_E_Comp}
\end{figure}

Comparing the prediction of a network trained on DS8max with the DFT results per individual simulations shows that the error is larger for higher energy configurations (see Fig. \ref{fig:ANI8_E_Comp}). The distribution of error shows exponential decay for small error region with a visibly fat tail. Further investigations also show that majority of the outlier configurations belong to a single molecule, hinting that careful error analysis beyond RMSE may be required for judging quality of network potentials. 

\clearpage
\subsection{Silicon}\label{ss:Silicon}
In this example a neural network potential is trained to reproduce the energies and forces of Silicon in solid and liquid phases.
While in the previous example the target output is obtained with DFT, here energy and forces are obtained using empirical Stillinger-Weber (SW) potential with the original parametrization~\cite{stillinger85}. Hence the target energy function that neural network is trained to approximate is indeed a simple function of atomic positions up to three body interactions.

The dataset consists of MD simulation snapshots at every $76.2$~fs after the equilibration period, in microcanonical (NVE) ensemble, for $216$ Si atoms in cubic box. The 90\% of the data is obtained with simulations at constant volume $V_0$, at a density corresponding to the liquid phase of Silicon at melting temperature,  where lattice parameter is ${16.053}$~\AA.
To sample the solid phase, atomic positions are initialized in cubic diamond phase and velocity of each atom is randomly chosen from a normal distribution that corresponds to a given initial instantaneous temperature. Approximately 100 independent MD simulations with equilibrium temperatures between 1~K and 2500~K are performed this way. 
To efficiently sample atomic environments corresponding to the liquid phase at different thermodynamic temperatures, first a set of atomic positions corresponding to the liquid radial distribution function is established via melting. This configuration is then used as the initial atomic configuration for approximately 100 MD simulations with equilibrium temperatures ranging between 1000~K and 5000~K. 

To sample the effect of lattice parameter on energy, additional molecular dynamics simulations at volumes equal to $\pm 10\%$ and $\pm 5\%$ of the previously fixed cell volume are performed with average equilibrium kinetic energies compatible with temperatures of 3~K, 300~K and 3000~K. Configurations from these MD simulations with various cell dimensions make up the remaining 10\% of the dataset.
The final dataset gathered in this fashion is composed of 10000 configurations and is split in 80\% and 20\% parts for training and validation purposes respectively.

Modified Behler-Parrinello type descriptors with radial and angular windows of $0.5-4.6$~\AA\ and $1.5-4.6$~\AA\ are generated. This cutoff radius is chosen conservatively larger than the SW interaction cutoff $3.8$~\AA, since the descriptor is only sensitive to the average bond length in an angle (see Eq.~\ref{eq:mBPang}). The descriptors included $15$ Gaussian centers for the radial part, and $4$ radial, $8$ angular centers for the angular part, resulting in descriptors of size $48$ for each atom. For the remaining parameters in equations \ref{eq:BPrad} and \ref{eq:mBPang}, the following choice is made: $\eta_{\rm rad}=16.0$~\AA$^{-2}$,  
$\eta_{\rm ang}=6.0$~\AA$^{-2}$,  
$\zeta=50$.  
In order to obtain the predicted forces analytically, derivative of the descriptors with respect to each atomic position is also calculated and stored. 

An all-to-all connected network with two hidden layers of sizes 32 and 16, both with Gaussian activation, is constructed. This network structure has $2113$ parameters which are optimized with Adam algorithm in order to minimize the sum of quadratic loss functions, Eq.~\ref{eq:quadloss} and Eq.~\ref{eq:Floss}, using energy and forces respectively. 

\begin{figure}[!hbt]
  \begin{center}
    \includegraphics[width=0.5\textwidth]{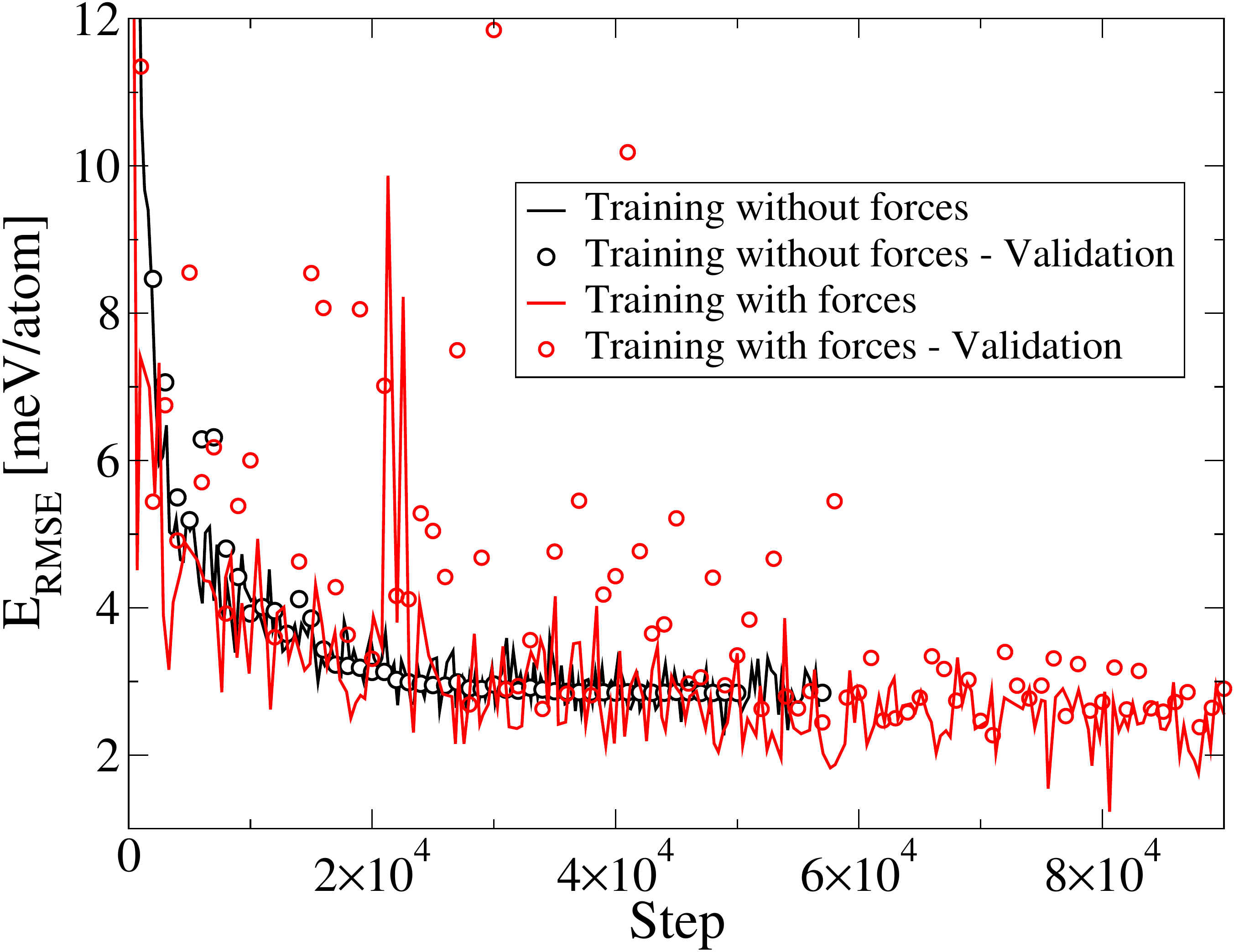}\\
    \includegraphics[width=0.5\textwidth]{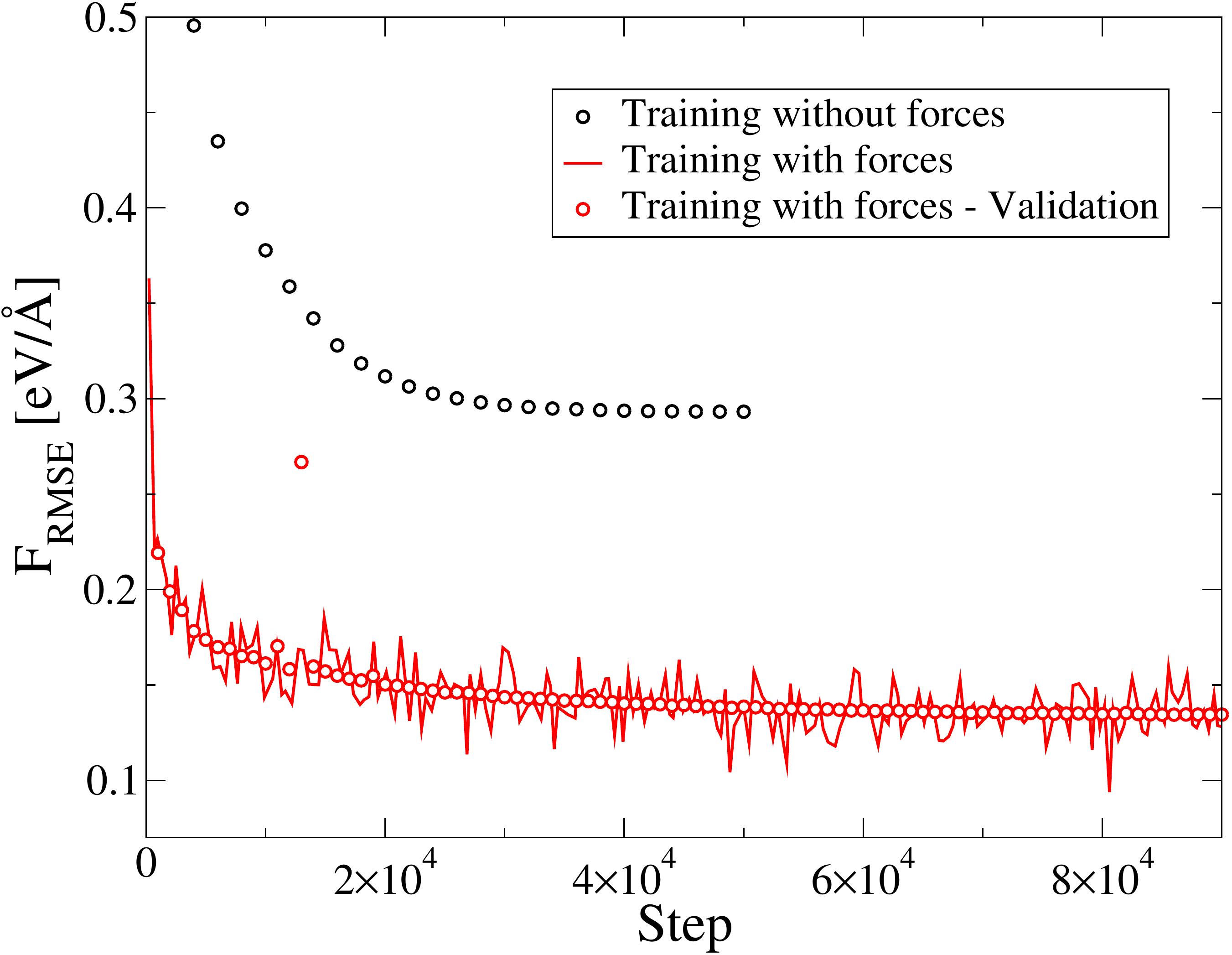}
  \end{center}
  \caption{{\bf Top:} RMSE of energy prediction on training batch (solid lines) and validation set (points). Training without loss from force predictions, $c_F=0$, uses a bigger batch size and quicker learning rate decay, resulting in a smoother training convergence behavior (black). The training converges more slowly when loss due to force prediction is considered with $c_F=1.3$ (red). The smaller batch size and slower decay contributes to slower convergence with increased stochasticity. {\bf Bottom:} RMSE of force prediction for each cartesian component on training batch (solid lines) and validation set (points). Note that early in the training, the total cost is dominated by the loss due to error on forces.}
  \label{fig:silicon_energy_evolution}
\end{figure}

First, the network parameters are trained using a loss function based on energy predictions alone, by setting the coefficient of the loss on forces, $c_F$, to zero. A batch size of $256$ simulations is used and the learning rate is decayed exponentially starting from $0.01$ with a decay rated of $5000$ steps, for $57000$ steps. The energy RMSE on the validation set is $2.9$~meV/atom 
with about $9\%$ of predictions having over $5$~meV/atom error, and with a maximum absolute error of $16.1$~meV/atom, in the same range of accuracy with results obtained modelling similar problems~\cite{BP07,Onat2018}. 
The force prediction of this network however is not accurate enough to perform reliable MD simulations, as the RMSE error on each force component calculated for the validations set is $0.295$~eV/\AA . The evolution of prediction error during training is summarized in Fig.~\ref{fig:silicon_energy_evolution}.

In order to better approximate the potential energy surface and obtain more accurate forces, training is performed with finite $c_F$. It is found that for the system size considered, values for $c_F \approx 1$ result in substantial reduction of error on force components. The result for $c_F=1.3$ is reported in the rest of this study. Since the inclusion of forces increases the required computation and memory for each step, a small batch size of $64$ simulations is chosen. The learning rate decay rate is increased to $22000$ steps to compensate for the smaller batch size and the additional loss. 

Using forces in guiding the parameter optimization dramatically increases the information about the target function the network is trained to approximate, hence lifts off possible degeneracy of network parameters. Yet, it can be seen in Fig.~\ref{fig:silicon_energy_evolution} that the accuracy on energy can be retained despite the penalty due to derivative matching, i.e. accurate energies and derivatives can be achieved simultaneously.
The RMSE error in energy prediction calculated for the validation set is $2.7$~meV/atom with about $9\%$ of predictions having over $5$~meV/atom error, and with a maximum absolute error of $12.9$~meV/atom. 
The RMSE for prediction of force components instead is $0.134$~eV/\AA, with only $4\%$ of the validation dataset predicted with error beyond $0.3$~eV/\AA, and with a maximum absolute error of $10.7$~eV/\AA. 
These prediction errors are similar to what is achieved in recent machine learning studies where forces were used in optimization of the machine learning parameters \cite{Bartok2018, Bonati2018}.

\begin{figure}[!hbt]
  \begin{center}
    \includegraphics[width=0.5\textwidth]{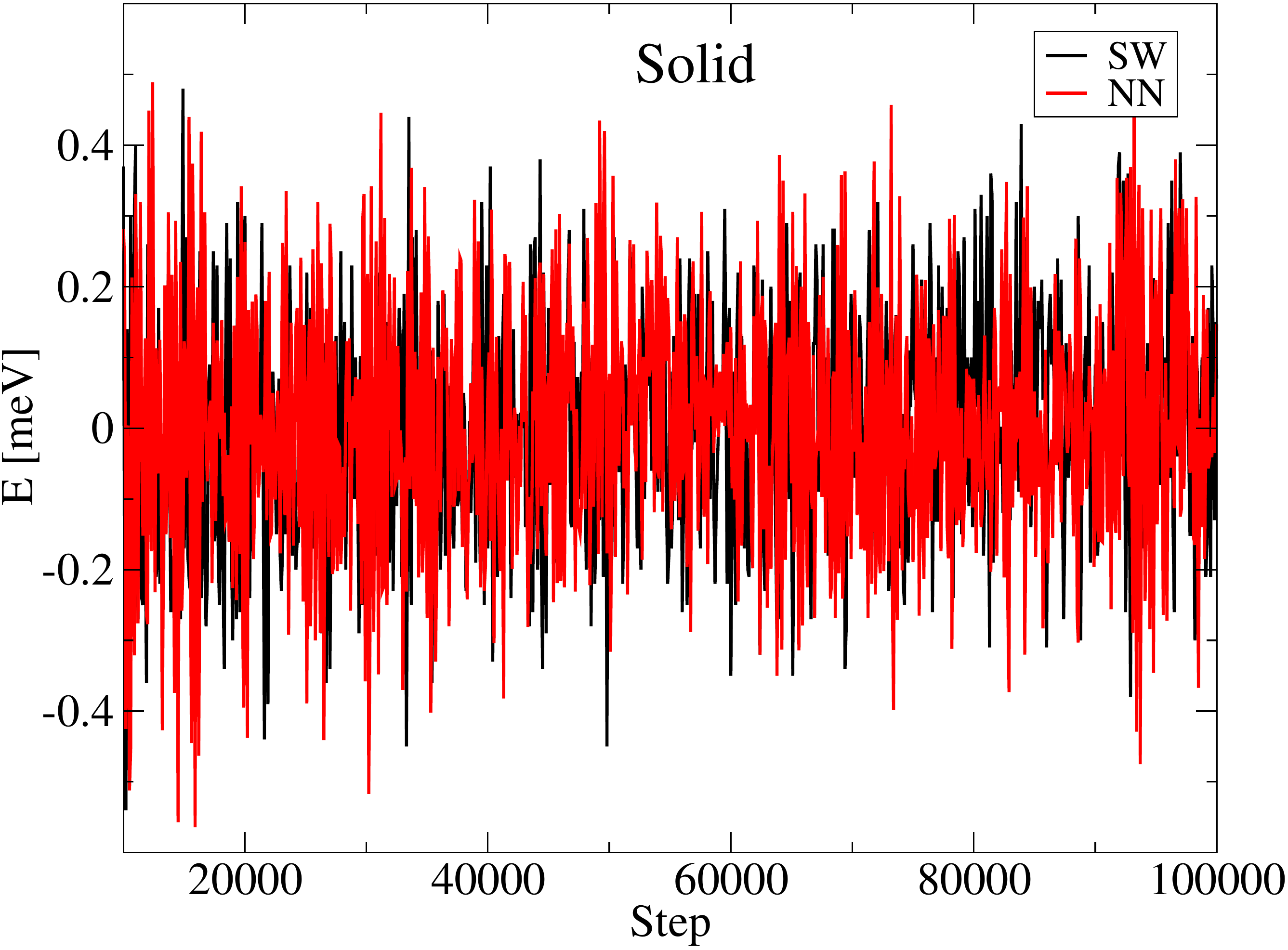}\\
    \includegraphics[width=0.5\textwidth]{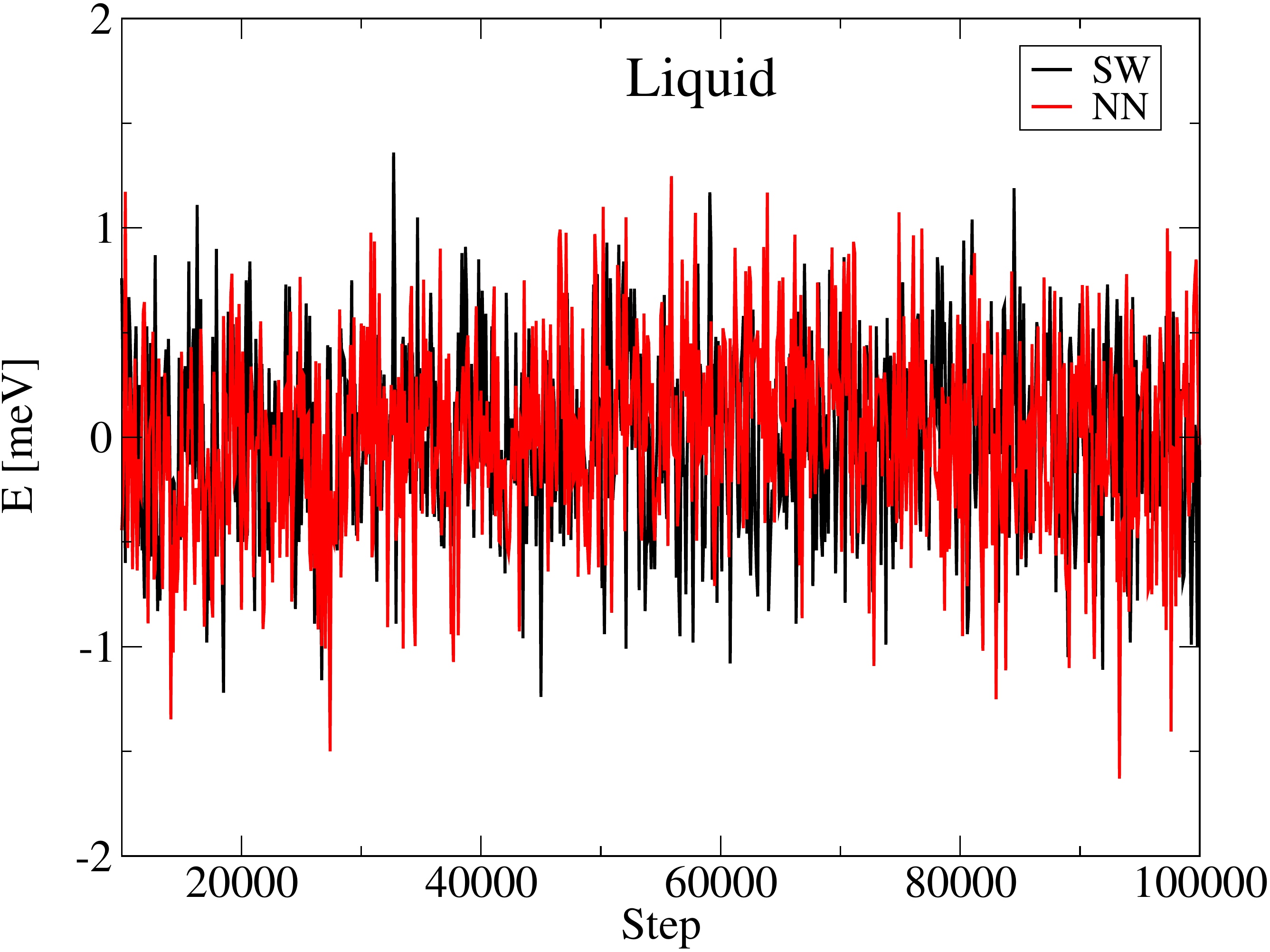}
  \end{center}
  \caption{Fluctuations in the total energy of the simulation of 216 Si atoms as a function of step during an NVE MD with a time step of 0.381~fs. {\bf Top:} The diamond phase at an average temperature of approximately $500$~K. Average energy difference of 77~meV between NN and SW potentials, $(E_{\rm NN} > E_{\rm SW})$, is removed for clarity. 
  {\bf Bottom:} Liquid phase at an average temperature of $1800$~K. Average energy difference of $954$~meV $(E_{\rm SW} > E_{\rm NN})$ is removed for clarity. This difference is considerably higher for liquid phase compared to the solid, indicating the increased discrepancy between the neural network prediction and SW potential for the high temperature phase. Despite this trend, conservation of energy is preserved due to analytically calculated forces.
  }
  \label{fig:si_solid_conservation_energy}
\end{figure}

Accurate forces that correspond to derivatives of energy enable conservation of total energy during MD simulations. An NVE simulation with average kinetic energy corresponding to approximately $500$~K is performed with time step $0.381$~fs for $38.1$~ps. It can be seen in Fig.~\ref{fig:si_solid_conservation_energy} that the conservation of energy is observed for the neural network potential to the same order of an equivalent SW simulation. When the average kinetic energy is increased to explore the liquid phase during the MD simulation, it is seen that the overall prediction error increases yet the conservation of energy is still retained \cite{G-angular-force-discontinuity}. 

\begin{figure}[!hbt]
  \begin{center}
    \includegraphics[width=0.5\textwidth]{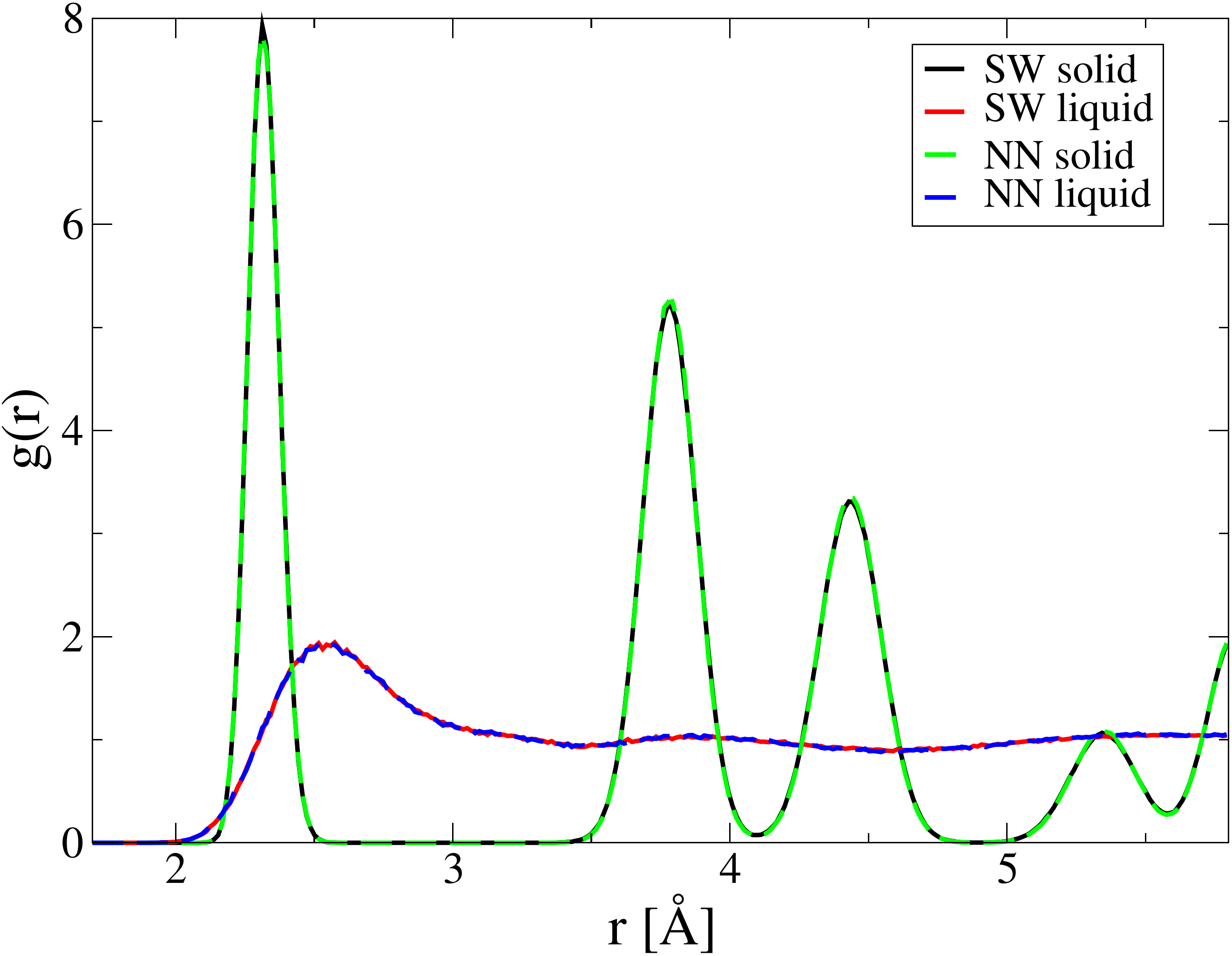}\\
  \end{center}
  \caption{Solid and liquid radial distribution function at at $500$~K and $2700$~K, respectively, computed during an NVE MD, for SW potential and the neural network potential. Although the accuracy in neural network energy prediction is lower for the high temperature phase as seen in Fig.~\ref{fig:si_solid_conservation_energy}, the radial distribution function is well reproduced.}
  \label{fig:si_gr}
\end{figure}

The pair correlation function observed during an NVE MD simulation is compared for solid ($500$~K) and liquid phases ($2700$~K) (See Fig.~\ref{fig:si_gr}). The prediction of the neural network potential matches the SW result well even for high temperature.

\begin{figure}[!hbt]
  \begin{center}
    \includegraphics[width=0.5\textwidth]{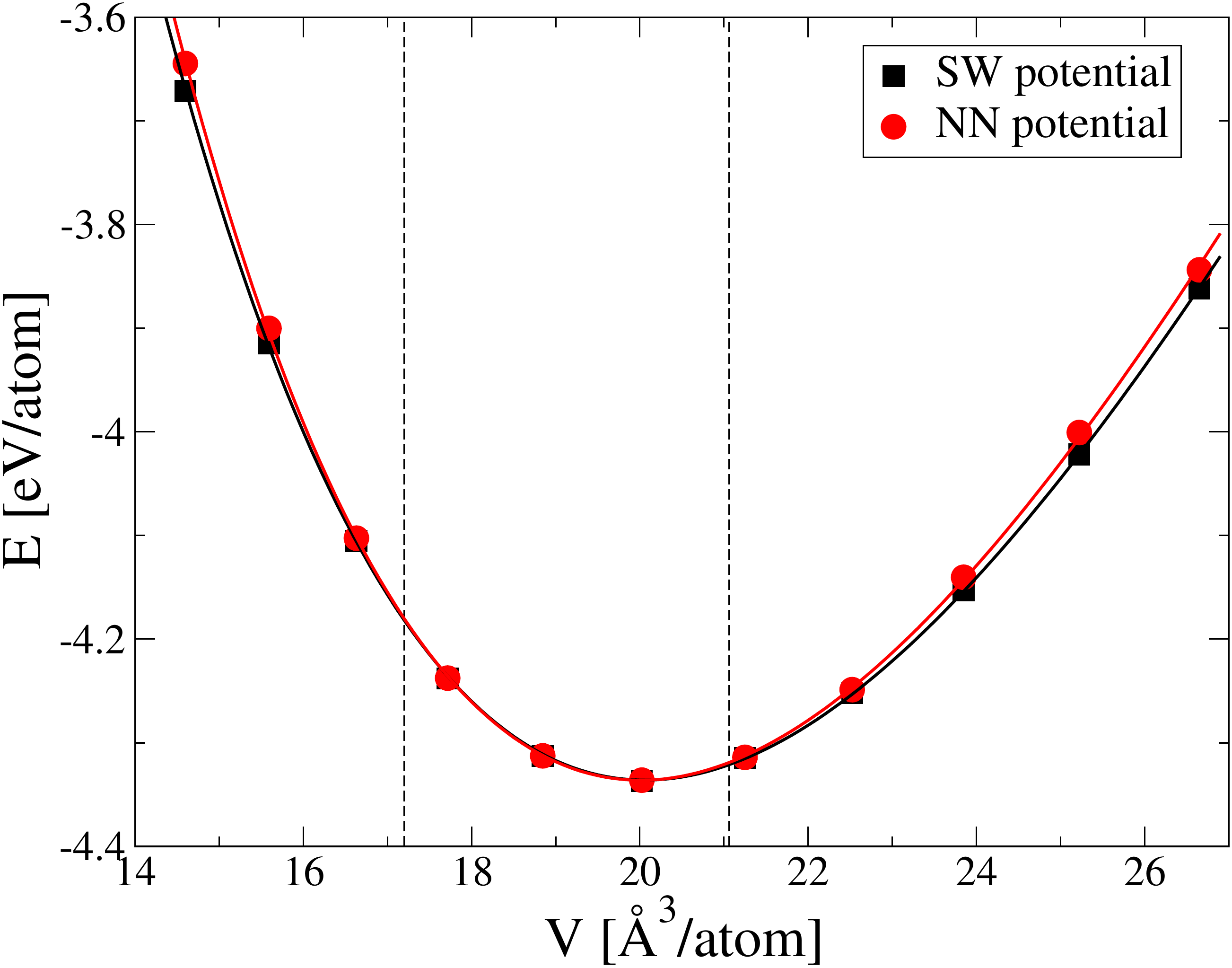}\\
  \end{center}
  \caption{Energy as a function of volume as obtained with the SW and the NN potentials. The lines are the Birch-Murnaghan equation of state fit to the data for each potential. The range of volume of the training data for the NN is shown with the dashed lines. The performance of the NN outside this range indicates its ability to extrapolate beyond the training set.}
  \label{fig:si_EV}
\end{figure}

Lastly, we assess the accuracy of predicted mechanical properties. Equation of state predicted with neural network potential shows good agreement with the one of SW for the volume range included in the training set, and it also performs reasonably well for the extended ranges (See Fig.~\ref{fig:si_EV}).
The bulk modulus obtained via Birch-Murnaghan equation of state fit is $105.5$~GPa and $101.4$~GPa in the case of neural network potential and SW respectively.
Further cell deformations lead to the estimate of the other elastic moduli, as reported in Table~\ref{tab:si_elastic_moduli}. The neural network potential reproduces the SW results within 2\% accuracy. 

\begin{table}[!hbt]
    \centering
    \begin{tabular}{|l|c|c|c|}
        \hline
        & NN  & SW & 
        \begin{tabular}[x]{c}
             Experiment \\
             (4.2~K)~\cite{Hall_1967} 
        \end{tabular} \\
        \hline
        $C_{11}$ & 148.4 & 151.4 & 167.5\\ 
        $C_{12}$ & 75.5 & 76.4 & 64.9 \\  
        $C_{44}$ & 56.8 & 56.4 & 80.2 \\ 
        \hline
    \end{tabular}
    \caption{Elastic constants in GPa for Silicon in diamond phase at $0$K calculated in this work for neural network potential (NN) and Stillinger-Weber potential (SW), and compared to experimental results from the literature. Note that network is trained to approximate the SW potential, which predicts $C_{12}$ and $C_{44}$ with limited accuracy with respect to the experiment.}
    \label{tab:si_elastic_moduli}
\end{table}


\section{Conclusion}
PANNA package offers a complete pipeline for the training of neural network potentials to be used in material science applications. It tackles preprocessing and management of external data from diverse sources, training and validation of neural network models based on them, and ultimately conversion of these models to a format compatible with multiple molecular dynamics codes.

PANNA offers flexibility in the definition of model architecture in terms of number of layers and their sizes, activation functions and regularizations. Considering the wide range of materials and structures considered in material science, such flexibility can be desirable. PANNA also offers the possibility to include the derivatives in the training, e.g. total energy and forces. In the simple example of neural network approximation of Stillinger-Weber potential, it is observed that aiming for the correct forces greatly improves the quality of the network model, while retaining the accuracy of its prediction for energy.

Thanks to the TensorFlow back end, PANNA can handle large networks and datasets, as demonstrated in the reproduction of state of the art training results from literature. At the same time, the simple input file interface to TensorFlow provides easy access to the features of this powerful neural network engine. 

Machine learning interatomic potentials is a rapidly growing field. Therefore, in time, addition of new methodologies, network architectures and descriptors are expected to take place as a part of maintenance of the package. As the user-base of the package increases, the input parsing and potential generating routines are planned to include a larger number of codes. In hosting an open source package, and enabling deposition of each model to an open online database, PANNA aims to provide a platform for users and developers that support the reproduciblity efforts in computational material science.

\section{Acknowledgements} 
The authors are thankful Alexie Kolpak, Asegun Henry and Spencer Wyant for an inspirational introduction to the universe of interatomic potentials and machine learning, 
to Stefano de Gironcoli for fruitful technical discussions, in particular for the smooth angular descriptor form that prevents discontinuous force derivative during dynamics,
to Ryan S. Elliott and Daniel Karls for their support with the KIM API,
to Qingjie Li for his contribution of \textit{ab initio} data parser.
E.K. is grateful for the financial support by European Union’s Horizon 2020 research and innovation program under grant agreement No. 676531 (project E-CAM). E.K. and F.P. acknowledge the MIT-FVG funds awarded in the 2017 MISTI (MIT International Science and Technology Initiatives) call by Regione Friuli Venezia Giulia that accelerated this work considerably. High performance calculations were carried out thanks to resources provided by CINECA and SISSA.


\bibliographystyle{elsarticle-num}
\bibliography{panna}

\begin{thebibliography}{10}
\bibitem{footnote1}
Currently, a patch for LAMMPS is distributed within PANNA package. It defines a
  new pair style that can read the weights computed with PANNA and use a NN
  force field for relaxations or molecular dynamics.
\bibitem{G-angular-force-discontinuity}
The angular descriptor for modified Behler-Parrinello functions have
  discontinuous derivative when three atoms are arranged collinearly which is
  found to give rise to violation of energy conservation during molecular
  dynamics. The angular descriptors are corrected with a smoother function
  where $\cos(\theta-\theta_s)$ term is replaced with $
  \cos(\theta)\cos(\theta_s)+\sqrt{1-\cos(\theta)^2+\epsilon\sin(\theta_s)^2}\sin(\theta_s)$
  and a normalization factor of
  $\frac{2}{\left[1+\sqrt{1+\epsilon\sin(\theta_s)^2}\right]} $ is introduced
  for MD simulations.
\end{thebibliography}

\end{document}